%%%%%%%%%%%%%%%%%%%%%%%%%%%%%%%%%%%%%%%%%%%%%%%%%%%%%%%%%%%%%%%%%
%%%%%%%%%%%%%%%%%%%%%%%       Paper1 : 
%%%        Generating-function method for tensor products
%%%
%%%%%%%%%%%%%%%%%%%%%%%%%%%%%%%%%%%%%%%%%%%%%%%%%%%%%%%%%%%%%%%%%  

\input harvmac.tex

%\input myharvmac.tex

%\draftmode

%=================================================================
% MACROS

%* on my mac I do not have these: so start by decommenting 
%* the following set and comment mine

 \input amssym.def   \def\Q{{\Bbb Q}}
  \def\N{{\Bbb N}}

%added this macro

%\input fonts \def\Z{{\blackB Z}} \def\R{{\blackB R}} \def\Q{{\blackB Q}}
%\def\C{{\blackB C}} \def\M{{\blackB M}} \def\N{{\blackB N}}

\def\non{\N}

\def\frac#1#2{{\textstyle{#1\over #2}}}

% standard tableaux
\def\b#1{\kern-0.25pt\vbox{\hrule height 0.2pt\hbox{\vrule
width 0.2pt \kern2pt\vbox{\kern2pt \hbox{#1}\kern2pt}\kern2pt\vrule
width 0.2pt}\hrule height 0.2pt}}
\def\ST#1{\matrix{\vbox{#1}}}
\def\STrow#1{\hbox{#1}\kern-1.35pt}
\def\bv{\b{\phantom{1}}}

% triangles
\def\tri#1#2#3#4#5#6#7#8#9{\matrix{#4\cr
	#3\quad#5\cr #2~\qquad #6\cr #1\quad #9\quad#8\quad#7\cr}}

%troisieme type de triangles

%triangles plus gros

%triangles plus gros#2

% alignements multiples
\def\eqalignD#1{
\vcenter{\openup1\jot\halign{
\hfil$\displaystyle{##}$~&
$\displaystyle{##}$\hfil~&
$\displaystyle{##}$\hfil\cr
#1}}
}
\def\eqalignT#1{
\vcenter{\openup1\jot\halign{
\hfil$\displaystyle{##}$~&
$\displaystyle{##}$\hfil~&
$\displaystyle{##}$\hfil~&
$\displaystyle{##}$\hfil\cr
#1}}
}

\def\text#1{\quad\hbox{#1}\quad}

\def\la{\lambda}

\def\nuh{{\hat \nu}}
\def\muh{{\hat \mu}}

\def\lah{{\hat \lambda}}

\def\om{{\omega}}
\def\y{{\infty}}

\def\x{{x^{-1}}}

\def\rw{\rightarrow}

\def\Om{\mathop{\Omega}\limits}
\def\O{{\displaystyle\Om_{\geq}}}
\def\Ox{{\displaystyle\Om_{\geq}^x}}
\def\Oy{{\displaystyle\Om_{\geq}^y}}

\def\Oxx{{\displaystyle\Om_{=}^x}}
\def\OR{{\displaystyle\Om_{=}^R}}
\def\OQ{{\displaystyle\Om_{=}^Q}}

\def\Nc{{\cal N}}

\def\Ec{{\cal E}}\def\Ac{{\cal A}}

\overfullrule=0pt

% Equations (overrides harvmac's equation macros)
\newcount\eqnum  
\eqnum=0
\def\eq{\eqno(\secsym\the\meqno)\global\advance\meqno by1}
\def\eqlabel#1{{\xdef#1{\secsym\the\meqno}}\eq }  

% References (overrides harvmac's reference macros)
\newwrite\refs 
\def\startreferences{
 \immediate\openout\refs=references
 \immediate\write\refs{\baselineskip=14pt \parindent=16pt \parskip=2pt}
}
\startreferences

\refno=0
\def\aref#1{\global\advance\refno by1
 \immediate\write\refs{\noexpand\item{\the\refno.}#1\hfil\par}}
\def\ref#1{\aref{#1}\the\refno}
\def\refname#1{\xdef#1{\the\refno}}

\newcount\exno
\exno=0
\def\Ex{\global\advance\exno by1{\noindent\sl Example \the\exno:

\nobreak\par\nobreak}}

\parskip=6pt
%===============================================================================

% PAGE TITRE
\Title{\vbox{\baselineskip12pt
\hbox{LAVAL-PHY-00-22}}}
%++ {\vbox {\centerline{Fusion-rule generating functions: I- The
%++ infinite-level limit}}}
%++ I think the new title is a little strange since
%++ we don't really talk about fusion rules. I'd
%++ prefer something simpler such as:
{\vbox {\centerline{Generating-function method for tensor products.}}}

%z changed superscript for me from 1 to 2
\centerline{L. B\'egin$^\natural$\foot{Work supported by NSERC
(Canada).}, C. Cummins$^{\sharp 2}$ and P. Mathieu$^\natural$\foot{Work supported
by NSERC (Canada) and FCAR (Qu\'ebec).} }
%z corrected my address
%\vskip.1in
\smallskip\centerline{$^\natural$ \it D\'epartement de Physique,
Universit\'e Laval, Qu\'ebec, Canada G1K 7P4}
\smallskip\centerline{$^\sharp$ \it CICMA, Department of Mathematics and Statistics, Concordia University,}
\centerline{Montr\'eal, Qu\'ebec, Canada H3G 1M8}
%\vskip .1in
%\centerline{\bf Abstract}
% \vskip .05in 
%\bigskip
\noindent
 
%

%+ rephrased somewhat:
%++ small changes

\noindent{\bf Abstract}: This is the first of two articles devoted to a
 exposition of the generating-function method for computing fusion
rules in affine Lie algebras.  The present paper is entirely devoted to
the study of the tensor-product (infinite-level) limit of fusions rules.
  We start by reviewing Sharp's
character method. An
alternative approach to the construction of tensor-product generating functions is
then presented which overcomes most of the technical difficulties associated
with  
the character method. It is based on the reformulation of the problem of
calculating tensor products in terms of the solution of a set of linear and
homogeneous Diophantine equations
whose elementary solutions represent ``elementary couplings''.  Grobner bases
provide  a tool for generating the complete set of relations between elementary
couplings and, most importantly, as an algorithm for specifying a  complete,
compatible  set
of ``forbidden couplings''. 

% possible to have complete but incompatible!!

\Date{11/98\ \ (revised 06/99,01/00)}

\let\n\noindent

%============================================================================== 

%+  I added 3 subsections here
\newsec{Introduction}

\subsec{Orientation}

%+ seems necessary to give a word on fusion; otherwise change the title
%+ but it seems to me that "tensor-product gen.fct" is not too appealing
%++ On the other hand "tensor-product generating functions" is
%++ the subject of the paper!

Fusion rules yield the
number of independent couplings between three given primary fields in conformal
field theories.  We are interested in fusion rules in unitary  conformal field
theories that have a Lie group symmetry, that is, those whose generating
spectrum algebra is an affine Lie algebra at integer level. These are the
Wess-Zumino-Witten models [\ref{V.G. Knizhnik and A.B. Zamolodchikov, Nucl. Phys.
{\bf B247} (1984) 83.},\ref{D. Gepner and E. Witten, Nucl. Phys. {\bf B278} (1986)
493.}\refname\GW]. Primary fields in these cases are in 1-1 correspondence with
the integrable representations of the appropriate affine Lie algebra at level
$k$.  Denote this set by $P_+^{(k)}$ and a primary field by the corresponding
affine weight $\lah$.  Fusion coefficients ${\Nc_{\lah\muh}^{(k)}}~^{\nuh}$ are
defined by the product
$$\lah\times \muh = \sum_{\nu\in P_+^{(k)}} {\Nc_{\lah\muh}^{(k)}}~^{\nuh}
\; \nuh\eq$$
(For a review of conformal field theory and in particular fusion rules,
see [\ref{P. Di Francesco, P. Mathieu, D. S\'en\'echal, {\it Conformal
Field Theory}, Springer Verlag 1997.}]; to a large extend we follow the
notation of this reference.)

In the infinite-level limit and for fields with finite conformal dimensions, the purely
affine condition on weight integrability is relaxed and the primary fields are 
solely characterised by their finite part, required to be an integrable weight of
the corresponding finite Lie algebra. Recall that a
%%## add "finite"
 finite weight ${\lambda}$  is
characterised  by its expansion coefficients in terms of the fundamental weights
${\omega}_{i}$
$${\lambda}=\sum_{i=1}^{r} \lambda_i
{\omega}_{i} = (\lambda_1,...,\lambda_r)\eq$$ where $r$ is the rank of the
algebra.  The numbers
$\la_i$'s are the Dynkin labels.  The set of weights with non-negative Dynkin
labels (the integrable weights) is denoted by
$P_+$.  

%((((
In the infinite-level limit the fusion coefficients reduce to
tensor-product coefficients:
$$\lim_{k \rightarrow \y} {\Nc_{\lah\muh}^{(k)}}~^{\nuh}= 
{\Nc_{\lambda\mu}}^{\nu}.
\eqlabel\limformula$$
where ${\Nc_{\lambda\mu}}^{\nu}$ is defined by
$$\la\otimes \mu = \sum_{\nu\in P_+} {\Nc_{\lambda\mu}}^{\nu}\;  \nu\eq$$
By abuse of notation, we use the same symbol for the highest weight and
the  highest-weight representation. Notice that
$$\Nc_{\lambda\mu}~^{\nu}=\Nc_{\lambda\mu \nu^*}\eq$$
where $\nu^*$
 denotes the highest weight of the representation
conjugate to that of $\nu$. Equivalently, $\Nc_{\lambda\mu \nu^*}$ gives the
multiplicity of the scalar representation in the triple product
$\la\otimes\mu\otimes\nu^*$.  

A tensor-product generating function codes the information for all the
tensor products of a given algebra in a single function
%((((
defined by 
$$G(L,M,N)= \sum_{\la,\mu,\nu \in P_+} {\Nc_{\lambda\mu}}^{\nu} L^\la M^\mu
N^\nu\eq$$
where
$L^\la=L_1^{\la_1}\cdots L_r^{\la_r}$ and similarly for
$M^\mu$ and $ N^\nu$.
%((((
$G$ can generally be expressed as a simple closed function of its
variables.  For instance, for $su(2)$, it reads
$$G(L,M,N)= {1\over (1-LM)(1-LN)(1-MN)}\eq$$

An example of basic global information that
can be deduced from  a generating function is the integrality as well as the
positivity of the tensor-product coefficients.  
More
 importantly, from our point of view, is that  in the context of fusion rules, the
construction of the simplest generating functions led to the discovery of the
notion of threshold levels [\ref{C.J. Cummins, P. Mathieu and M.A. Walton, Phys.
Lett. {\bf B254} (1991) 390.}\refname\CMW].  Moreover, as  shown in the sequel
paper, setting up a fusion generating function is a way to obtain explicit
expressions for these threshold levels.  Our new approach to fusion-rule
generating functions, which originates from the generalisation of techniques
developed in the present paper on tensor products, leads to a further new
concept, that of a fusion basis.

%+ More to say on the interest for gen. fct.? agree with this? 

%---------------------------------------------------------------------------
\subsec{Overview of the paper}

%" add references to sections

 The present article is organised as follows.  We start by explaining
  in detail the construction of tensor-product generating
functions for finite Lie algebras. The first construction  which
is presented is the character method developed by Sharp and his collaborators 
(section 2).

Although it is conceptually very simple,  the
character method is limited by its inherent computational difficulties:
the disproportion between the simplicity of the resulting form of the generating
function and the intermediate calculations is enormous. This motivates our
alternative approach to the construction of
tensor-product generating function. 
It is based on the reformulation of the problem of calculating
tensor products in terms of the solution of a set of linear and homogeneous
Diophantine equations (cf. section 3).   The elementary
solutions of these Diophantine equations represent  ``elementary couplings''. 
For $sp(4)$, the use of the
Berenstein-Zelevinsky inequalities to obtain the elementary couplings and their
relations (cf. the analysis of section 6) is new.

The key difficulty is finding
the numerous relations that exist in general between the
elementary solutions. 
{}From the Diophantine-equation point of view, the decomposition of a solution may
not be unique because different sums of elementary solutions could yield the same
result. 
To solve this problem
we first ``exponentiate'' it: given a solution $\alpha =
(\alpha_1,\dots,\alpha_k)$ to our system of linear Diophantine
equations, we introduce formal variables $X_1,\dots,X_k$
and consider the monomial $X_1^{\alpha_1} \dots X_k^{\alpha_k}$.
The linear span, $R$,
of all such monomials is a ``model'' for the
generating function for the solutions to the original
set of linear Diophantine equations (see section 5),
since the Poincar\'e series of $R$ is 
the required generating function. 
This series can be calculated using 
Grobner basis methods.

 For $su(N)$ there is  a remarkable
graphical construction for computing tensor product
multiplicities, the famous
Berenstein-Zelevinsky triangles.  These are introduced in section 6.
We also
discuss the analogous construction for $sp(4)$, whose diagrammatic representation
is new. But the main interest of these re-formulations is that it yields a
%pedestrian and systematic way of obtaining the elementary couplings from the
%## I think that "pedestrian" has a very negative implication. It would
%## mean that it is too simple to be of interest.
%ý ok
simple and systematic way of obtaining the elementary couplings from the
construction of a vector basis.  
Thus we get a new way of constructing the
%((((
corresponding
generating functions.

%  In Appendix A and B, we present graphical constructions that follows
% from the equality version of the Diophantine inequalities for the
% $su(N)$ and the
% $sp(4)$ cases respectively. For $su(N)$, these are the famous
% Berenstein-Zelevinsky triangles, while for $sp(4)$, the diagrams are new. In
% Appendix C, we present a vector basis approach to the construction of
% tensor-product generating functions.

%------------------------------------------------------------------

%============================================================================== 

\newsec{Generating-function for tensor products: the character method}

%------------------------------------------------------------------

\subsec{The character method for the construction of the tensor-product
generating function: the 
$su(2)$ case}

The method developed by Sharp and collaborators for constructing generating
functions for tensor products is based on manipulations of the character
generating functions [\ref{R.
Gaskell, A. Peccia and R.T. Sharp, J. Math. Phys. {\bf 19} (1978)
727; J. Patera and R.T. Sharp, in {\it Recent advances in group theory and
their applications to spectroscopy}, ed. J. Domini, New-York, Plenum; J. Patera
and R.T. Sharp, in Lecture Notes in Physics (New York, Springer Verlag, 1979)
vol. 94, p. 175;  M.
Couture and R.T. Sharp, J. Phys. {\bf A13} (1980) 1925; R.T. Gaskell and R.T.
Sharp, J. Math. Phys. {\bf 22} (1981) 2736;  C. Bodine and R.T. Gaskell, J. Math.
Phys. {\bf 23} (1982) 2217;   R.V. Moody, 
J. Patera and R.T. Sharp, J. Math. Phys. {\bf 24} (1983) 2387; 
J. Patera and R.T. Sharp, J. Phys. {\bf A13} (1983) 397;  Y. Giroux, M. Couture and R..T. Sharp, J. Phys. {\bf
A17} (1984) 715.}\refname\SP]. Although
simple in principle, these manipulations become rather cumbersome as the rank of
the algebra is increased. To illustrate the method, we will work in complete
detail the simplest example, the
$su(2)$ case.

%# La liste complete conprend aussi les refs 
% suivantes mais on les utilise plus bas 
%  R.T Sharp and D. Lee, Revista Mexicana de
% Fisica {\bf 20} (1971)
% 203; M. Hongoh, R.T. Sharp and D.E. Tilley, J. Math. Phys. {\bf
% 15} (1974) 782;  R.T. Sharp, J. Van der Jeught and 
% J.W.B. Hughes, J. Math. Phys.
% {\bf 26} (1985) 901;

The first step is the derivation of the character generating function.
The
Weyl character formula for a general algebra of rank $r$ and a  highest-weight
representation
$\lambda$ is
$$\chi_\lambda= {\xi_{\lambda+\rho}\over \xi_\rho}\eq$$ where $\rho$ is the
finite Weyl vector, $\rho= \sum_{i=1}^r\om_i$,
and where the characteristic function $\xi$ is defined as
$$\xi_{\la+\rho}=\sum_{w\in W}\, \epsilon(w)e^{w(\la+\rho)}\eq$$
where $\epsilon(w)$ is the signature of the Weyl reflection $w$ and $W$ is
the Weyl group.  

For $su(2)$, $W$ contains two elements: $1,s_1$.  With $$x=
e^{\om_1}\eq$$ the 
$su(2)$ characteristic function $\xi$ for the representation of highest weight
$m\om_1\equiv(m)$ is
$$x^{m+1}-x^{-m-1}\eq$$
The character reads then
%ýý add an intermediate equality
$$\chi_m = { x^{m+1}-x^{-m-1}\over  x-x^{-1}}={ x^{m}-x^{-m-2}\over 
1-x^{-2}}=  x^m+x^{m-2}+\cdots+ x^{-m}\eqlabel\aa$$ The character generating
function
$\chi_L$ is obtained by multiplying the above expression by $L^m$ where $L$ is a
dummy variable, and summing over all positive values of
$m$: 
$$\eqalign{ \chi_L(x) &= \sum_0^\y L^m\chi_m = {1\over x-x^{-1}} \sum_0^\y
L^m(x^{m+1}-x^{-m-1})\cr &= {1\over 1-x^{-2}}\left({1\over 1-Lx}-{x^{-2}\over
1-Lx^{-1}}\right) = {1\over (1-Lx) (1-Lx^{-1})}\cr}\eq$$
We should point out here that in all generating functions in 
this paper, expressions of the form $1/(1-a)$ should be formally expanded
in positive powers of $a$. So for example, $1/(1-Lx^{-1})= 1 +
Lx^{-1} + L^2x^{-2} + \dots$.
By construction, the character of the highest weight $(m)$ can be
recovered from the power expansion of $\chi_L$ as the coefficient of the term
$L^m$. The characteristic generating function $\xi_L$ is defined by
$$\chi_L(x) = {\xi_L\over \xi_0}\eq$$
and it reads
$$ \xi_L(x) = {x-x^{-1}\over (1-Lx) (1-Lx^{-1})}={x\over 1-Lx}-{x^{-1}\over
1-Lx^{-1}}\eq$$
the last form being the one that results directly  from (\aa).  

The tensor product of two highest-weight representations can be obtained from
the product of the corresponding characters:
%* add the factor {\Nc_{mn}}^\ell:
$$\chi_m \chi_n= \sum_\ell  {\Nc_{mn}}^\ell\;\chi_\ell\eq$$  
This information can be extracted
from the product of the corresponding generating functions.  We are thus led to
consider the product $\chi_L (x)\chi_M(x)$.  To simplify the analysis of the
resulting expression, notice that the information concerning the
representations occurring in the tensor product is coded in the leading term of
the character, i.e., the term $x^{m+1}$.  
%((((
%The rest of the representation  is
%easily reconstructed by the application of the Weyl group and the action of the
%ladder operator.  Actually, 
To insure that every positive
power of $x$ singles out a highest-weight representation, we can multiply both sides by
$\xi_0$.   To read off these terms, we can focus on the terms
with strictly positive powers of
$x$ in the product 
$\chi_L(x)
\chi_M (x)\xi_0(x)$.
%((((
If we require the Dynkin label of the representations (and
not their shifted value), it is  more convenient to divide  by $x$ before doing
the projection, now restricted to the non-negative powers of $x$.  The
truncation of an expression by its negative powers of $x$ will be denoted by the
MacMahon symbol [\ref{P. MacMahon, {\it Combinatory analysis}, 2 vols
(1917,1918), reprinted by Chelsea, third edition, 1984.}\refname\Mac]
$\Omega$, defined by
$$\Ox \,\sum_{-\y}^\y c_nx^n = \sum_{n\geq0} c_nx^n\eq$$
When there is no ambiguity concerning the variable in terms of which the
projection is defined, it is omitted from the $\Omega$ symbol.

We are thus interested in the projection of the following expression
$$\eqalign{ \chi_L(x) \chi_M(x) \xi_0 (x)\x &=  \chi_L(x)
\xi_M(x) \x \cr &= {1\over (1-Lx) (1-Lx^{-1})}\left({1\over
1-Mx}-{x^{-2}\over 1-Mx^{-1}}\right)\cr}
\eqlabel\twopiece$$ For these manipulations, we use systematically the following
simple identities:
$$\eqalign{ {1\over (1-Ax) (1-Bx^{-1})} &= {1\over (1-AB)} \left({1\over
1-Ax} + {Bx^{-1}\over 1-Bx^{-1}}\right)\cr &=  {1\over (1-AB)} \left({Ax\over
1-Ax} + {1\over 1-Bx^{-1}}\right)\cr
&=  {1\over (1-AB)} \left({1\over
1-Ax} + {1\over 1-Bx^{-1}} -1\right)\cr}\eq$$
%ýý add an equality above and a comment below

%((((
There are two terms to analyse.  The first is
$${1\over (1-Lx) (1-Lx^{-1})(1-Mx)}= 
{1\over (1-Lx) (1-LM) }\left( {1\over 1-Mx} +{Lx^{-1}\over 1-Lx^{-1}} \right)\eq$$
%(((
The first part is not affected by the projection and the second can be
written as
$${Lx^{-1}\over (1-Lx) (1-LM)(1-Lx^{-1}) } = {Lx^{-1}\over
(1-LM)(1-L^2)}\left({Lx\over 1-Lx}+ {1\over 1-L\x}\right)\eq$$
The second term of this expression contains only negative powers of $x$ and
%(((
can thus be ignored and the first part is unaffected by the projection. We have thus,
for the first term of (\twopiece)
$$\O\, {1\over (1-Lx) (1-Lx^{-1})(1-Mx)} =  {1\over (1-Lx) (1-LM) }\left( {1\over
1-Mx}+{L^2\over1-L^2}\right)\eqlabel\prem$$
%((((
The projection of the second term of (\twopiece) is:
$$\eqalign{& \O\, {x^{-2}\over (1-Lx) (1-Lx^{-1})(1-Mx^{-1}) }\cr
 \qquad\qquad \quad &= \O\, {x^{-2}\over 
(1-Lx^{-1})(1-LM)}\left( {1\over 1-Lx}+{M\x \over 1-Mx^{-1} }\right)\cr
\qquad \qquad \quad &= \O\, {x^{-2}\over 
(1-Lx^{-1})(1-LM) (1-Lx) }\cr
\qquad\qquad \quad &= \O\, {x^{-2}\over 
(1-LM)(1-L^2) }\left( {Lx\over 1-Lx}+{1\over 1-L\x}\right)\cr
\qquad\qquad \quad &= \O\, {Lx^{-1}\over 
(1-LM)(1-L^2)(1-Lx)}\cr
\qquad \qquad \quad &= {Lx^{-1}\over 
(1-LM)(1-L^2)}\left( {1\over 1-Lx} -1\right) \cr
 \qquad\qquad \quad &=  {L^2\over 
(1-LM)(1-L^2)(1-Lx)}\cr}  \eqlabel\seco$$
Subtracting (\seco) from (\prem), we find that 
$$\O\, \chi_L(x) \xi_M(x)\,\x\, = {1\over 
(1-LM)(1-Lx)(1-Mx)}\eq$$
Replacing $x$ by $N$, we thus get
$$G^{su(2)}(L,M,N)= {1\over 
(1-LM)(1-LN)(1-MN)}\eqlabel\sufun$$

%* introducing a vague but simple notion
%*  of elementary coupling before the abstract setting: 
%*{\it 
%*Notice that the building
%*blocks of the generating function are the composites $LM,
%*\, LN, $ and $MN$. These composite building blocks are called the elementary
%*couplings.   Every coupling can be decomposed as a product of these elementary couplings.
%*This notion will be made more precise in the following subsection. }

%In the generating function, the terms that appear in the denominators are the
%{\it elementary couplings}.  If there are {\it syzygies}, namely, relation between
%the elementary couplings, they appear in the numerator. So for this simple case,
%there are three elementary couplings:
%$$\eqalignD{ &E_1= LM: \quad &(1)\otimes(1)\supset(0), \cr & E_2=
%LN: &(1)\otimes(0)\supset(1),\cr & E_3= MN: &(0)\otimes(1)\supset(1)\cr}
%\eqlabel\elesud$$ and no syzygy.
%-----------------------------------------------------------------

\subsec{The abstract setting: Poincar\'e series, elementary couplings and
relations; defining a model}

 As we shall see it is frequently useful
have a {\it model},
$R$,  for a generating function $G(X_1,\dots,X_k)$ such as (\sufun). 
By this we mean
a commutative $\Q$-algebra with an identity, graded by $\non^k$, (
$\non = \{0,1,2,3,\dots\}$) 
$$R=\oplus_{\alpha\in \non^k} R_\alpha\, , \qquad \quad R_\alpha R_\beta\subseteq
R_{\alpha+\beta}\eq$$  
%z changed wording a little to make clear what properties we
%z require
%z with Hilbert series:
and such that its  Poincar\'e series
%%## minor change
(also frequently called Hilbert series)
$$
F(R) = \sum_{\alpha\in \non^k} \dim_\Q(R_\alpha) X^\alpha 
$$
satisfies
$$
F(R)  = G(X_1,\dots,X_k).\eq
$$
%* Here I replace A,B,C with E_1, E_2 E_3 and define X_i 
%* in terms of L,M,N: ok?
For example, for (\sufun), with $X_1= L, \, X_2= M, \, X_3=N$,  we can take
$R=\Q[E_1,E_2,E_3]$, which is the polynomial ring generated by
the formal variables $E_1,E_2,E_3$ (in fact
all our examples $R$ is either a subring or quotient of a polynomial ring) with
the grading of $E_1,E_2$ and $E_3$ being
$(1,1,0)$, $(1,0,1)$ and $(0,1,1)$. The homogeneous subspaces
are spanned by $E_1^a E_2^b E_3^c$, $a,b,c\in \non$ with grade
$(a+b,a+c,b+c)$ and so
$$
F(R) = \sum_{(a,b,c)\in\non^3} X_1^{a+b}X_2^{a+c}X_3^{b+c}
= G^{su(2)}(X_1,X_2,X_3)
\eq$$
as required.

If $R$ is generated by elements $E_1,\dots,E_s$
and is a model for a generating function $G$ for tensor products
(or fusion products)
 then we
call $E_1,\dots,E_s$ a set of ``elementary couplings''
for $G$.

% (in the form $X_1^{a_1}X_2^{a_2}\cdots$ where the grading of $X_i$ is the
%natural one, that is $(0,\cdots 0,1,0,\cdots,0)$ with the 1 in the $i$-th
%position) 
%? above: add the natural grading.
%+ I see that it is no more necessary to add this (but you mention 
%+ latter "natural grading") 

%+ replace ".." by it for grading and model variables
It should perhaps be stressed that {\it a priori}
the  variables 
$X_1,\dots,X_k$ and
$E_1,\dots,E_s$ are unrelated. 
We shall refer to the $E$'s as {\it model variables}
and the $X$'s as {\it grading variables}.
If the grading vector of $E_i$ is $\alpha^i$ , $i=1,\dots,s$
then 
there is an associated monomial in 
the grading variables: $X^{\alpha^i}$, for which we
will use the notation $g(E_i)$.
For example in 
the above example we have $g(E_1) = X_1^1X_2^1X_3^0 = LM$.
However, to avoid tedious repetition when writing down
generating functions we shall often write, for example,
$1/(1-E_1)$ rather than $1/(1-g(E_1))$. In {\it all} such
cases where model variables appear in a generating function
they should be replaced by the corresponding monomial
in the grading variables. 

In the  case of tensor products we use
the notation ``$E : g(E) : \hbox{\rm product}$''
to denote a set of elementary couplings with their ``exponentiated'' grading 
and the corresponding term in 
 the tensor product. So in the example above we would write:
 $$\eqalignD{ &E_1: LM: \quad &(1)\otimes(1)\supset(0), \cr & E_2:
 LN: &(1)\otimes(0)\supset(1),\cr & E_3: MN: &(0)\otimes(1)\supset(1)\cr}
 \eqlabel\elesud$$ 

Having made the distinction  between
grading and model variables, it should be noted that 
there are 
cases where we can identify the model as a ring 
generated by monomials in the grading variables.
So in the above example we could {\it define}
$E_1=LM$, $E_2=LN$ and $E_3=MN$ and take the
model for our generating function to be the
subring of $Q[L,M,N]$ generated by $E_1,E_2$ and
$E_3$. However, it is not always desirable, or
even possible, to make this identification. 

We close
this
section with two examples of how models for the $su(2)$ character
generating function can be constructed.
%+ seems that it is better to use su(2) instead of SU(2) above
%+ uniform notation: algebra: su(2) and group : SU(2)

The first method, which has been exploited by Sharp et al (see
[\SP]) to
construct character generating functions, amounts to finding an algebra $R$ which
is a module for the Lie algebra $su(2)$ and such that, as
an $su(2)$ module, $R$ is isomorphic to $\oplus_{i\geq 1} V_i$ where
$V_i$ is the irreducible $su(2)$ module of dimension $i$.

In this case we can take $R=\Q[p,q]$ with the generators of
$su(2)$ being given by differential operators: 
$$
h=p{{\partial}\over{\partial p}} -q{{\partial}\over{\partial q}},\quad
x_- = q{{\partial}\over{\partial p}},\quad
x_+ = p{{\partial}\over{\partial q}}\eq
$$
The $su(2)$ highest-weight vectors are $p^i$, $i\geq 0$ and a basis
of the irreducible submodule of dimension $i$ is just given by
the monomials of degree $i$ in $p$ and $q$. We can give $R$ an  $\non^3$ grading
by taking the degree of $p$ to be $(1,1,0)$ and of $q$ to be $(1,0,1)$.
%+ I would like to add a word to explain the 3 indices:
Here the first grading index specifies the representation while the other two
refer to a particular weight. As $R=\Q[p,q]$ the Poincar\'e function for $R$
is, 
$${1\over {(1-p)(1-q)}}$$
with the understanding, as explained above,
that $p$ and $q$ should be replaced
by the corresponding expression in terms of the grading variables.
%+ Why not calling these grading variables X_1,X_2,X_3?
%+ because it is not natural: here is
%+ a reformulation that makes natural our choice: ok?
Let us denote these grading variables here by $L$ (which exponentiates the
representation index) and $x,y$ (exponentially related to the weights).  The
Poincar\'e function reads then
$${1\over {(1-Lx)(1-Ly)}}\eqlabel\poin$$

Another way of constructing a model for the weight generating function, 
 which
makes more natural the $\non^3$ grading,  is to observe that the complete set
$SU(2)$ weight vectors  of
finite dimensional irreducible $su(2)$  modules 
are in 1-1 correspondence
with one-rowed Young tableaux. If the Young  tableau has $c$ boxes
filled with $a$ 1's and $b$ 2's then there is a constraint
$$a+b-c=0,\quad a,b,c\geq 0\eqlabel\lind$$
and so the solutions to this linear
Diophantine equation are in 1-1 correspondence with the complete
set of  $SU(2)$ weight vectors. Thus to find a model for the
weight generating function it is sufficient to find a model
for the solutions to (\lind).
It is not difficult to see that every solution to this equation
is a linear combination (with non-negative coefficients)
of the two fundamental solutions:
$(a,b,c) = (1,0,1)$ and $(a,b,c) = (0,1,1)$. Let $R$ be the
subring of $\Q[A,B,C]$ generated by the monomials $E_1=AC,\, E_2=BC$.
Considering the exponents of the monomials $E_1$ and $E_2$, we
see that the monomials in $R$ correspond to the solutions of 
(\lind) and hence taking the 
%+ this is where you mention:
natural grading on $R$
ensures that the Poincar\'e series of $R$ is the generating
function for the solutions to  (\lind) and hence is the required
generating function. In this example there are no relations between
$E_1$ and $E_2$ and so $R$ is isomorphic to the polynomial ring 
in
two variables (as expected) and so the Poincar\'e function is once
again (with $A\rw x, B\rw y, C\rw L$) given by (\poin).
%%## here I have refereed to 2.24 instead of repeating the eq.
%
%+ again: why dont we keep ABC: to make contact with the above : so
%+ we should give the correspondence (simple but makes reading easier
% $${1\over {(1-Lx)(1-Ly)}}\eq$$

%------------------------------------------------------------------

\subsec{Multiple $su(2)$ tensor products}

In order to illustrate  the occurrence of 
%%## add a segment:
linear relations between elementary couplings,
consider the problem of finding the multiplicity of a
given representation $\zeta$ in the triple product
$\la\otimes \mu\otimes \nu$. In terms of character generating functions, this
amounts to considering the product
$\chi_L(x)\chi_M(x)\chi_N(x) \supset \chi_P(x)$,
or equivalently, $\chi_L(x)\chi_M(x)\xi_N(x)\x \supset \xi_P(x)\x$. The left
side is
then projected onto positive powers of $x$. We are thus led to consider
 $$\O\; {1\over (1-Lx)(1-L\x)(1-Mx)(1-M\x)}\left({1\over 1-Nx}- {x^{-2}\over
1-N\x}\right)\eq$$
The projection of each term is worked out as previously and the
resulting expression is found to be, with $x$ replaced by $P$:
$$G(L,M,N,P) = {1-LMNP\over (1-LP)(1-MP)(1-NP)(1-LM)(1-LN)(1-MN)}\eqlabel\ffa$$
This is the sought for generating function. Here we would like
to have a model with 6 elementary couplings corresponding
to the terms in the 
%* add: denominator
denominator of the generating function:
$$\eqalignD{
&E_1: LM: \quad &(1)\otimes(1)\otimes(0)\supset(0)\cr
&E_2: LN:  &(1)\otimes(0)\otimes(1)\supset(0)\cr
&E_3: LP:  &(1)\otimes(0)\otimes(0)\supset(1)\cr
&E_4: MN:  &(0)\otimes(1)\otimes(1)\supset(0)\cr
&E_5: MP:  &(0)\otimes(1)\otimes(0)\supset(1)\cr
&E_6: NP:  &(0)\otimes(0)\otimes(1)\supset(1)\cr}\eq$$
%%## modify the insertion of the footnote
and there must be a linear relation (in
this context, such a relation is often called a {\it syzygy} in the physics
literature - see in particular [\SP] and related works)  between the following
products 
%* add:
(signalled by a term in the numerator)  
which has grading $LMNP$:
%* footnote added
$$E_1E_6, \qquad E_2E_5, \qquad E_3E_4\eq$$

It is not difficult to see that a model is given by
$\Q[e_1,e_2,e_3,e_4,e_5,e_6]/I$ where $E_i=e_i+I, i=1,\dots,6$
and $I=(ae_1e_6 + be_2e_5 +c
e_3e_4)$ is the ideal generated by the polynomial $ae_1e_6 +
be_2e_5 +ce_3e_4$ for any choice of  $a,b,c\in\Q$ not all zero.

The elements of $R$ have the form $m + I$ with $m\in
\Q[e_1,\dots,e_6]$. However there is no canonical way
of choosing the representatives $m$. Take for example
the case $a=b=c=1$. (Usually we will construct a model 
for our generating function as explained above and this
construction will fix the values of $a,b$ and $c$).
 In $R$ we have $ E_1E_6 = -(E_2E_5 + 
E_3E_4)$ and so we can take as a basis for $R$ the
set of (equivalences classes of ) monomials which do not contain the product $E_1E_6$. In this case we say that we have chosen to make
$E_1E_6$ a `forbidden product'.
Similarly we can forbid the products $E_2E_5$ or
$E_3E_4$. As we shall see later, the choice of forbidden products
corresponds to a choice of {\it term ordering}.

%# change the word `fusion' for `composition'
Before leaving this example, we would like to rework it from a different
point of view, as an illustration of the `composition' technique of generating
functions.
 Let $G(L,M,R)$ describe the tensor product corresponding to
$\chi_L\chi_M\supset\chi_R$ and similarly let $G(Q,N,P)$ correspond to
$\chi_Q\chi_N\supset\chi_P$. We are interested  the product
$\chi_L(x)\chi_M(x)\chi_N(x) \supset \chi_P(x)$, but treated from the product
of the two generating functions $G$.  We thus want to enforce the constraint $R=Q$
in the product 
$G(L,M,R)G(Q,N,P)$. 
%# Add a foot: Chris : do you have a more precise ref?
%##Not really, I know about it from Bob. He said that Stanley was
%##the person who told him and others about MacMahon. But who did what
%##and when I don't know.
%ý let us leave it that way then
The idea -- which is used in the references in [\SP] mainly in
relation with the construction of generating functions for branching functions
-- is to multiply this product by
$(1-Q^{-1}R^{-1})^{-1}$ and, in the expansion in powers of $R$ and $Q$, keep only
terms of order zero in both variables: with an obvious notation we have
$$ \eqalign{\OR\OQ\, & G(L,M,R)G(Q,N,P) {1\over1-Q^{-1}R^{-1}}\cr
  &\qquad\,=\, \OR\OQ\,\sum_n A_n(L,M)\, R^n \sum_m B_m(N,P)\,Q^m\sum_\ell
R^{-\ell}Q^{-\ell}\cr&\qquad\,= \sum_p A_p(L,M) B_p(L,M)\cr}\eq$$
which is manifestly equivalent to considering
$$ \Oxx\,G(L,M,x)G(\x,N,P) \eq$$
With the explicit expressions for the generating functions, we have thus
$$\Oxx\, {1\over (1-Lx)(1-Mx)(1-LM)} {1\over (1-P\x)(1-N\x)(1-NP)}\eq$$ A brief
 and by now standard analysis yields directly the generating function (\ffa). 

%-------------------------------------------------------

\subsec{The $sp(4)$ case}

%z specific equation number inserted. Is this the right equation?
%" Yes
% As a final example, consider the $sp(4)$ case.  With the $x_i$ defined as before,
As a final example, consider the $sp(4)$ case.  
With the $x_i= e^{\om_i},  i=1,2$,
the characteristic function is found to be 
$$\eqalign { \xi_{(m,n)}
&=x_1^{m+1}x_2^{n+1}-x_1^{-m-1}x_2^{m+n+2}-x_1^{n+m+5}x_2^{-n-1}
+x_1^{m+2n+3}x_2^{-m-n-2}\cr
 &+x_1^{-m-2n-3}x_2^{n+m+2}-x_1^{m+1}x_2^{-m-n-2}
-x_1^{-m-2n-3}x_2^{n}+x_1^{-m-1}x_2^{-n-1}\cr}\eq$$
and the characteristic generating function is
$$\eqalign{ \xi_{L_1,L_2}& = {1\over (1-L_1x_1) (1-L_1x_1x^{-1}_2) (1-L_2x^{-1}_2)
(1-L_2x_1^{-2}x_2)}\cr
&\times\left( {1+L_2\over (1-L_2x_1^2x^{-1}_2)(1-L_2x^{-1}_2)}+
{(1+L_2)L_1x_1\over (1-L_1x_1)(1-L_2x_1^2x^{-1}_2)} \right.\cr
& \left. \qquad\qquad
 +{L_1x^{-1}_1x_2\over
(1-L_1x_1)(1-L_1x^{-1}_1x_2)}\right)\cr}\eq$$
{}From this  we construct the character
generating function and then we can proceed to the tensor-product generating
function.  This is again extremely cumbersome. The result is 
[\ref{M. Hongoh, R.T. Sharp and D.E. Tilley, J.Math. Phys. {\bf 15}
(1974) 782.}\refname\Hongo]
$$\eqalign{ &G^{sp(4)}(L_1,L_2,M_1,M_2,N_1,N_2)\cr &=[(1-M_1 N_1) (1-L_1 N_1)
(1-L_1 M_1) (1-M_2 N_2) (1-L_2 N_2) (1-L_2 M_2)]^{-1}
\cr&\times 
\left( {1\over(1-L_2 M_1
N_1) (1-L_2 M_1^2 N_2)}+{L_2 M_2 N_1^2 \over(1-L_2 M_1 N_1) (1-L_2 M_2 N_1^2)}
\right. \cr 
&\left.+{L_1^3 M_2^2 N_1 N_2 \over (1-L_1 M_2 N_1) (1-L_1^2 M_2 N_2)} 
 +{L_1 M_2 N_1 \over (1-L_1 M_2 N_1) (1-L_2 M_2 N_1^2)}\right. \cr
 &\left. +{L_1^2 M_2 N_2 \over (1-L_1 M_1 N_2) (1-L_1^2 M_2 N_2)} +
{L_1 M_1 N_2
\over (1-L_1 M_1 N_2) (1-L_2 M_1^2 N_2)}\right) \cr}\eq$$
{}From this expression, we read off the following list of  elementary couplings
(recall that the first variable is a model variable and then we write
the corresponding monomial in the grading variables):
$$\eqalign{
A_1 : M_1 N_1,\quad\quad\quad & A_2 : L_1 N_1,\quad\quad\quad ~~A_3 : L_1 M_1 \cr
B_1 : M_2 N_2,\quad\quad\quad & B_2 : L_2 N_2,\quad\quad\quad ~~B_3 : L_2 M_2 \cr
C_1 : L_2 M_1 N_1,\quad\quad & C_2 : L_1 M_2 N_1,\quad\quad ~C_3 : L_1 M_1 N_2 \cr
D_1 : L_1^2 M_2 N_2,\quad\quad & D_2 : L_2 M_1^2 N_2,\quad\quad D_3 : L_2 M_2
N_1^2 . \cr} \eq $$ 
However, not all the products of the model variables can be
linearly independent: there are linear
relations between:
$$\eqalignT{ &C_i C_j, \qquad &A_k D_k, \quad &{\rm and}\quad  A_i A_j B_k\cr
&D_i D_j,  \qquad &A_k^2 B_i B_j,  \quad &{\rm and}\quad  B_kC_k^2\cr
& C_i D_i,  \qquad & A_j B_k C_k,  \quad &{\rm and}\quad  A_k
B_j C_j\cr}\eqlabel\sprel$$ for $i,j,k$ a cyclic permutation of $1,2,3$ and
repeated indices are not summed. (It is plain  
that the three sets of products found to be linearly related must
have the same Dynkin labels.)  A specific form of the  generating function,
%%## add:
as expressed in terms of the elementary couplings, 
amounts to a specific choice of
a set of forbidden couplings among those that are related by a linear relation. 
%%## Q:should we mention the question of the compatibility
%%## of the choices of forbidden couplings?

%==================================================================
\newsec{Tensor-product descriptions }

%------------------------------------------------------------------

\subsec{The need for a tensor-product description}

%z minor wording changes
It is clear  that one major technical complication of the character method is that
it starts at  too fundamental a level, namely the character of the separate
representations.  One natural way to proceed is to start from a combinatorial description of the
tensor-product rules. Such a description already
takes into account the action of the Weyl group and encodes the various
subtractions of the singular vectors. 

But how do we make the connection with the generating-function approach?
The key is to find a combinatorial description which can be expressed
as a set of linear Diophantine inequalities. 
Given this set of inequalities, 
%Given a basis, 
%there is a well-defined way to get to the generating function. Actually, in this
there is an algorithm, again due to 
%ýý add a foot + ref
%z fixed typos, changed order of footnotes and punctuation
MacMahon, for
constructing a generating function.  (This is an adaptation of a
method developed by Elliot [\ref{E.B. Elliott, Quart. J. Math. {\bf 34} (1903)
388.}] for the analysis of linear Diophantine equalities and for this reason
the algorithm is often referred to as the Elliot-MacMahon method. 
%%## add:
For a detailed discussion of the algorithm, 
see in particular vol. 2
section VIII of [\Mac].)  This method is  conceptually similar to the character
method, except that the starting point is substantially closer to the end result. 
See section 7.3 for a slight generalisation of this algorithm.
% Before we pursue our line of reasoning, 

Although the description of tensor products via linear Diophantine
equations is a more efficient route to finding the generating function
than the character one,
complications associated to the
$\Omega$ projections remain a source of technical difficulty that severely limits the
practical applicability of the method.

%As a by-product of the character method, provided the generating
%function is sufficiently simple, we can obtain a model and hence
%define elementary couplings and
%the forbidden couplings. The MacMahon's method was originally conceived as
%a technique to generate the elementary couplings and their linear relations.
%On the other hand, from the knowledge of these data, that is, the elementary
%couplings and the relations, it is rather simple in principle to construct the
%generating function. 
A more powerful approach to our problem is 
to use the techniques of computational algebra. We start with a description
of the tensor-product multiplicities as solutions to linear Diophantine
inequalities. 
Efficient algorithms exists for finding the fundamental solutions
to these inequalities 
[\ref{G. Huet,
{\it An algorithm to generate the basis of solutions to 
homogeneous Diophantine equations}, Information Processing Lett.
{\bf 7} (1978) 144-7.}\refname\Huet]. 
{}From these we find directly a model for the generating
function using Grobner basis techniques. (This is roughly the
%* add:
%(or equivalently the set of elementary couplings -- which correspond to
% the elementary solutions of the
%Diophantine equations --  and a choice of forbidden couplings).
inverse of MacMahon's method which was originally conceived as a technique to
generate the elementary couplings and their linear relations through the construction
of the generating function.  Here, the elementary couplings and their relations are
first obtained and used as the input for the construction of the generating function.)

%thus to
%invert the procedure: start from the tensor-product basis and generate the
%elementary couplings and the precise forms of the syzygies and then construct
%the generating function.  

% A little difficulty remains toward the full implementation of this program, which
%is  that when there are many syzygies, there is a compatibility criterion in the
%choice of forbidden couplings that needs to be taken into account.  This issue
%is addressed in the following section.

%============================================================================== 

\newsec{The LR rule ($su(N)$)}

For $su(N)$ tensor products there is a particularly
convenient description based on Littlewood-Richardson tableaux
supplemented by the stretched-product operation (defined below) 
[\ref{M.Couture,
C.J.Cummins and R.T.Sharp, J.Phys {\bf A23} (1990) 1929.}\refname\CCS].

Integrable weights in $su(N)$ can be represented by tableaux: the weight
$(\la_1,\la_2,
\cdots ,\la_{N-1})$ is associated to a left justified tableau of $N-1$ rows 
%(((( I changed the expression for the second row. This
%(((( is correct isn't it?
with $\la_1+\la_2+\cdots +\la_{N-1}$ boxes in the first row, $\la_2+\la_3+\cdots
+\la_{N-1}$ boxes in the second row, etc.  Equivalently, the tableau has $\la_1$
columns of 1 box, $\la_2$ columns of 2 boxes, etc. The scalar representation has
no boxes, or equivalently, any number of columns of $N$ boxes. 

The  Littlewood-Richardson
rule is a simple combinatorial description of the tensor product of two $su(N)$
representations $\la\otimes \mu$.   The second tableau ($\mu$) is filled with
numbers as follows: the first row with
$1$'s, the second row with $2$'s, etc. All the boxes with a $1$ are then added  to
the first tableau according to 
following restrictions:

\n 1) the resulting tableau must be regular: the number of 
boxes in a
given row must be smaller or equal to the number of boxes in the row 
immediately above;

\n 2) the resulting tableau must not contain two boxes 
marked by $1$
in the same column.

\n All the boxes marked by a $2$ are the added to
the resulting tableaux according to the above two rules (with $1$
is replaced by $2$) and the further restriction:

\n 3) in counting from right to left and top to bottom, the 
number of
$1$'s must always be greater or equal to the number of $2$'s.

\n The process is repeated with the boxes marked by a $3, 4, \cdots, N-1$, with
the additional
rule that the number of
$i$'s must always be greater or equal to the number of $i+1$'s when counted from
right to left and top to bottom.
 The resulting Littlewood-Richardson (LR) tableaux are the Young
tableaux of the irreducible representations occurring in the decomposition.

%*% add a ref to CCS and modify the 1st eq: min(k,N-1)->k and k<j.
These rules can be rephrased in an algebraic way as follows [\CCS]. Define
$n_{ij}$ to be the number of boxes $i$ that appear in the LR tableau in the row
$j$. The LR conditions read
$$\lambda_{j-1}+\sum_{i=1}^{k-1} n_{i,j-1}-\sum_{i=1 }^{k} n_{ij}\geq 0
\quad\quad\quad 1\leq k < j\leq N   \eqlabel\nijrang$$
and
$$\sum_{j=i}^{k} n_{i-1 \, j-1}-\sum_{j=i}^{k} n_{ij} \geq
0 \quad\quad\quad 2\leq i \leq k \leq N \quad {\rm and} \quad i\leq N-1.
\eqlabel\nijlr$$
The weight $\mu$ of the second tableau and the weight $\nu$ of the resulting LR
tableau are respectively given by 
$$\eqalign{  
  \sum_{j=i}^{N} n_{ij} & =\sum_{j=i}^{N-1} \mu_{j} \quad\quad\quad\quad
i=1,2,..., N-1 \quad ,\cr 
\nu_j -\lambda_j+\sum_{i=1}^{N-1} n_{i\, j+1} & =\sum_{i=1}^{{\rm
min}(j,N-1)} n_{ij} \quad\quad\quad\quad j=1,2,..., N-1 ~. \cr} 
\eqlabel\nijtrois$$ 
Hence, given three weights $\la,\mu$ and $\nu$, the number of 
%%## change "positive" for "non-negative"
non-negative
integers solutions $\{n_{ij}\}$ satisfying the above conditions gives the
multiplicity 
$\Nc_{\lambda
\mu}^{\quad\nu}$ of $\nu$ in the tensor product $\la\otimes \mu$.

The combined equations (\nijrang) and (\nijlr) constitute a set of linear and
homogeneous inequalities.  
%z insert ref to Stanley
As described in [\ref{R.P. Stanley, Duke Math. J. {\bf 40} (1973) 607; 
{\it Combinatorics and Commutative Algebra}, (Boston:
Birkhauser) (1983).}\refname\Stan], 
the Hilbert basis theorem guarantees that every solution can be
expanded in terms of the elementary solutions of these inequalities.

We can construct a model for the
solutions of the equations (\nijrang) and (\nijlr) by introducing
new formal variables $A_i$, $1\leq i\leq t$ where $t$ is the total
number of variables in (\nijrang) and (\nijlr). Then the subring
of $\Q[A_i; 1\leq i\leq t]$ generated by the monomials $A^\alpha$
with $\alpha$ a solution of (\nijrang) and (\nijlr) provides the
required model. This ring $R$ will be generated by a finite set of
monomials $E_j$ $1\leq j\leq s$ which we call elementary couplings
corresponding to the elementary solutions of (\nijrang)
and (\nijlr). Thus $R$ is isomorphic to $\Q[e_1,\dots,e_s]/I$ 
under the mapping $\phi: e_i\rightarrow E_i$ where $I$
%((((
is some ideal. Each element of $I$ corresponds, via the map $\phi$,
to a relation between the elementary couplings.  

In the case of LR tableaux, there is a nice pictorial representation
of the model $R$. Consider the set of formal linear combinations
of 
%+ here and below I replace Tableaux by LR tableaux: agree ?
LR tableaux with rational
coefficients. It is given a ring structure by  
defining the {\it stretched product} of two LR tableaux (denoted by $\cdot$) to
be the tableau obtained by
fusing the two tableaux and reordering the numbers in each row in increasing order
[\CCS].  More algebraically, if we denote the empty boxes of a LR tableau by a 0,
so that
$$ n_{0j} 
=\sum_{i=j}^{N-1} \lambda_i \quad\quad\quad\quad j=1,2, ...,N-1\eq$$
we can characterise  completely a tableau by the data $\{n_{ij}\}$ with now
$i\geq 0$. 
It is clear the set of numbers $\{n_{ij}\}$ with $i\geq 0$, or equivalently,
$\{\la_i, n_{ij}\}$ with $i\geq1$, is a complete set of variables for the
description of the tensor products. 
Then, the tableau obtained by the stretched product of the
tableaux  $\{n_{ij}\}$ and  $\{ n'_{ij}\}$ is simply described by the numbers 
$\{n_{ij}+n'_{ij}\}$. 
Here is a simple example:
$$\matrix{\ST{\STrow{\bv\bv\b1}\STrow{\b1\b1\b2}\STrow{\b2\b3}\STrow{\b4}
}\cr} \cdot \matrix{\ST{\STrow{\bv\bv\b1}\STrow{\bv\b1\b2}\STrow{\bv\b2}
}\cr}  =  
\matrix{\ST{\STrow{\bv\bv\bv\bv\b1\b1}\STrow{\bv\b1\b1\b1\b2\b2}
\STrow{\bv\b2\b2\b3}\STrow{\b4} 
 }\cr}\quad  \eq $$

This ring of tableaux is isomorphic to the model $R$ constructed
above and we do not distinguish between them. Thus we specify a set of
elementary couplings (i.e. a set of generators of $R$) as a set
of elementary LR Tableaux.
%+ above I replace Young Tableaux by LR tableaux

%------------------------------------------------------------------

\subsec{Example: the $su(2)$ case}

The complete set of inequalities for $su(2)$ variables $\{\la_1, n_{11},
n_{12}\}$ is simply 
$$\la_1 \geq n_{12} \qquad
 n_{11}\geq 0 \qquad n_{12}\geq 0\eqlabel\inedeux$$
The other weights are fixed by $$ \mu_1 =n_{11}+n_{12}\qquad 
\nu_1=\la_1+n_{11}-n_{12}\eq$$
By inspection, the elementary solutions of this set of inequalities are
$$(\la_1, n_{11}, n_{12}) = (1,0,1), \quad (1,0,0), \quad (0,1,0)\eq$$
which correspond respectively to $E_1, E_2, E_3$ in (\elesud). 
These correspond to the
following LR tableaux:
$$E_1: \quad \ST{\STrow{\bv}\STrow{\b1}}\, ,  \qquad E_2: \quad \ST{\STrow{\bv}}\,
,
\qquad E_3: \quad \ST{\STrow{\b1}}\eqlabel\tabdu$$It is also 
manifest
that there are no linear relations between these couplings.  The generating
function is thus simply:
$$G^{su(2)}= {1\over (1- E_1)(1- E_2)(1- E_3)}\eq$$

%------------------------------------------------------------------

\subsec{Example: multiple tensor products in the  $su(2)$ case}

Consider the problem of finding 
the multiplicity of the representation $\zeta$ in the triple
product
$\la\otimes \mu\otimes \nu\supset \zeta$. As a first step, the LR rule applies
as before: with $n_{11}+n_{12}= \mu_1$, we have
$\la_1\geq n_{12}$. After the first product, we re-apply the LR rule with now
$\la_1$ replaced by $\la_1+n_{11}-n_{12}$ and $n_{ij}$ replaced by $m_{ij}$ with
$m_{11}+m_{12}= \nu_1$.  The LR gives $\la_1+n_{11}-n_{12}\geq m_{12}$. The two
inequalities for the $su(2)$  quadruple product are
then:
$$\la_1\geq n_{12}\qquad \la_1+n_{11}-n_{12}\geq m_{12}\qquad
n_{ij}\geq 0\qquad m_{ij}\geq 0\eqlabel\multiin$$ The elementary solutions are
then, in the order: name of the coupling, corresponding Dynkin labels and the
5-vector
$(\la_1,n_{11},n_{12},m_{11},m_{12})$,:
$$\eqalignT{
 &E_1: &(1)\otimes(1)\otimes(0)\supset (0)\qquad& (1,0,1,0,0)\cr  
 &E_2: &(1)\otimes(0)\otimes(1)\supset (0)\qquad &(1,0,0,0,1)\cr 
 &E_3: &(1)\otimes(0)\otimes(0)\supset (1)\qquad &(1,0,0,0,0)\cr 
 &E_4: &(0)\otimes(1)\otimes(1)\supset (0)\qquad &(0,1,0,0,1)\cr 
 &E_5: &(0)\otimes(1)\otimes(0)\supset (1)\qquad &(0,1,0,0,0)\cr 
 &E_6: &(0)\otimes(0)\otimes(1)\supset (1)\qquad &(0,0,0,1,0)\cr}\eqlabel\mutipo$$
The linear relation, whose existence was signalled by the character method, is
$$E_3E_4= E_2E_5: (1,1,0,0,1), \qquad \not= E_1E_6: (1,0,1,1,0)\eq$$
 Choosing to forbid the product $E_3E_4$, the
generating function can be written in the form
$$\eqalign{ G &= {1-E_3E_4\over
(1-E_1)(1-E_2)(1-E_3)(1-E_4)(1-E_5)(1-E_6)}\cr
&= \left(\prod_{i=1,2,5,6}{1\over 1-E_i}\right)\left({1\over 1-E_3}+{E_4\over
1-E_4}\right)\cr}\eqlabel\multi$$
The latter form makes manifest the absence of $E_3 E_4$.

We could represent the  elementary couplings in terms of tableaux, where the boxes
with 1's refers to the $\mu$ tableau and those with 2's originate from the $\nu$
tableau.  (Warning: the resulting tableaux describing the four-products are not
%z changed wording  a little
 necessarily LR tableaux.) Hence, $n_{1j}$ gives the number of 1's in row $j$ of the
composed tableau while $m_{1k}$ gives the number of 2's in row $k$.  The elementary
tableaux are 
$$\eqalignT{ &E_1:\quad \ST{\STrow{\bv}\STrow{\b1}}  \qquad &E_2: \quad
\ST{\STrow{\bv}\STrow{\b2}}\, ,
& E_3: \quad \ST{\STrow{\bv}}\cr
&E_4:\quad \ST{\STrow{\b1}\STrow{\b2}}  \qquad &E_5: \quad
\ST{\STrow{\b1}}\, ,
& E_6: \quad \ST{\STrow{\b2}}\cr}\eq$$
{}From this representation, the relation reads
$$E_3E_4= E_2E_5: \ST{\STrow{\bv\b1}\STrow{\b2}}, \qquad \not= E_1E_6:
\ST{\STrow{\bv\b2}\STrow{\b1}}\eq$$

%------------------------------------------------------------------

\subsec{Example: the $su(4)$ case}

%* add one ineq here

The $su(4)$ LR conditions are:
$$\eqalignD{
&\la_1 \geq n_{12}\qquad &n_{11} \geq n_{22}\cr
&\la_2 \geq n_{13}\qquad & n_{11}+n_{12} \geq n_{22}+ n_{23}\cr
&\la_2+n_{12} \geq n_{13}+n_{23}\qquad &n_{11}+n_{12}+n_{13} \geq n_{22}+
n_{23}+n_{24}\cr
 &\la_3 \geq n_{14}\qquad &n_{22} \geq n_{33}\cr
&\la_3+n_{13} \geq n_{14}+n_{24}\qquad &n_{22}+n_{23} \geq n_{33}+ n_{34}\cr
&\la_3+n_{13}+n_{23} \geq n_{14}+n_{24}+  n_{34}\qquad ~\cr
}\eqlabel\inequatre$$

The tensor-product elementary couplings are:
$$\eqalign{
A_1 &:\ST{\STrow{\b1}\STrow{\b2}\STrow{\b3} }, \quad  
 A_2:\ST{\STrow{\bv}\STrow{\bv}\STrow{\bv}\STrow{\b1}}, \quad
 A_3:\ST{\STrow{\bv}},\quad
B_1 :\ST{\STrow{\b1}\STrow{\b2} }, \quad  
 B_2:\ST{\STrow{\bv}\STrow{\bv}\STrow{\b1}\STrow{\b2}}, \quad
 B_3:\ST{\STrow{\bv}\STrow{\bv}},\cr
C_1 &:\ST{\STrow{\b1} }, \quad  
 C_2:\ST{\STrow{\bv}\STrow{\b1}\STrow{\b2}\STrow{\b3}}, \quad
 C_3:\ST{\STrow{\bv}\STrow{\bv}\STrow{\bv}},\quad
D'_1 :\ST{\STrow{\bv}\STrow{\bv}\STrow{\b1} }, \quad  
 D'_2:\ST{\STrow{\bv}\STrow{\b1}}, \quad
 D'_3:\ST{\STrow{\bv}\STrow{\b1}\STrow{\b2}},\cr}\eqlabel\elequ$$
together with
$$
D_1 :\ST{\STrow{\bv\b1}\STrow{\bv}\STrow{\b2}\STrow{\b3} }, \, 
D_2 :\ST{\STrow{\bv\b1}\STrow{\bv\b2}\STrow{\bv}\STrow{\b3} }, \, 
D_3 :\ST{\STrow{\bv\b1}\STrow{\bv}\STrow{\bv}\STrow{\b2} },\, 
E_1 :\ST{\STrow{\bv\bv}\STrow{\bv\b1}\STrow{\bv}\STrow{\b2} }, \, 
E_2 :\ST{\STrow{\bv\b1}\STrow{\bv}\STrow{\b2} }, \, 
E_3 :\ST{\STrow{\bv\b1}\STrow{\bv\b2}\STrow{\b1}\STrow{\b3} }\eqlabel\elequu$$
The Dynkin-label transcription of the elementary couplings reads
$$\eqalignD{
&A_1:(0,0,0)\otimes(0,0,1)\supset(0,0,1)\qquad
&D_1':(0,1,0)\otimes(1,0,0)\supset(0,0,1)\cr
&A_2:(0,0,1)\otimes(1,0,0)\supset(0,0,0)\qquad
&D_2':(1,0,0)\otimes(1,0,0)\supset(0,1,0)\cr
&A_3:(1,0,0)\otimes(0,0,0)\supset(1,0,0) \qquad
&D_3':(1,0,0)\otimes(0,1,0)\supset(0,0,1) \cr 
&B_1:(0,0,0)\otimes(0,1,0)\supset(0,1,0)\qquad
&D_1:(0,1,0)\otimes(0,0,1)\supset(1,0,0)\cr 
&B_2:(0,1,0)\otimes(0,1,0)\supset(0,0,0)\qquad 
&D_2:(0,0,1)\otimes(0,0,1)\supset(0,1,0)\cr 
&B_3:(0,1,0)\otimes(0,0,0)\supset(0,1,0)\qquad
&D_3:(0,0,1)\otimes(0,1,0)\supset(1,0,0) \cr 
&C_1:(0,0,0)\otimes(1,0,0)\supset(1,0,0)\qquad 
&E_1:(1,0,1)\otimes(0,1,0)\supset(0,1,0)\cr 
&C_2:(1,0,0)\otimes(0,0,1)\supset(0,0,0)	\qquad
&E_2:(0,1,0)\otimes(0,1,0)\supset(1,0,1)\cr  
&C_3:(0,0,1)\otimes(0,0,0)\supset(0,0,1)\qquad
&E_3:(0,1,0)\otimes(1,0,1)\supset(0,1,0)\cr}\eq$$   
For $su(4)$ there are 15
relations [\ref{R.T Sharp and D. Lee, Revista Mexicana de Fisica {\bf 20} (1971)
203.}\refname\SL,\CCS] :
$$ \eqalignT{ &D_j^{'}  D_ k = C_i E_i \qquad &D_j D_k^{'}  = B_i C_j C_k  \qquad
&E_i E_j  = B_k D_k D_k^{'} \cr
&D_i E_i  = C_j B_k D_k \qquad
&D_i^{'}E_i   = B_j D_j^{'} C_k \qquad &\quad {} \cr}
\eqlabel\mmmhb$$
with $ i,j,k$ a cyclic permutation of $1,2,3$.

%! Chris: check the following
%!  statement: the problem related to non compatible selection...
%+ I understand that all is ok now

 To construct the generating function, we need to select forbidden couplings. 
It turns out that when there are more that one relation, complications
may arise.  We must ensure that the selected forbidden
couplings are complete, which means that no further (usually higher-order)
relations are required for a unique decomposition of a given coupling. 
 A technique that is
tailor-made for dealing with problems of that type is that of Grobner bases. 
This will be introduced in the next section. At this point, we simply indicate a
complete choice of forbidden couplings, namely   
%" the given set $\{C_iC_j, D_iD_j, C_iD_j\}$ ref to sp(4): corrected 
$\{E_iE_j, D'_iE_i, D_iE_i, D_jD'_i,D'_jD_i\}$.  This yields then a model for
the generating function, which then reads [\SL,\CCS] :
$$\eqalign{
G^{su(4)}=& (
\prod_{i=1}^3~ \tilde{A}_i \tilde{B}_i \tilde{C}_i ) (
\tilde{D}_1' \tilde{D}_2' \tilde{D}_3'
+E_1 \tilde{E_1} \tilde{D}_2' \tilde{D}_3'
+D_3 \tilde{D}_3 \tilde{D}_3'\tilde{E}_1 \cr 
&+D_2 \tilde{D}_2 \tilde{D}_3 \tilde{E}_1
+D_1 \tilde{D}_1 \tilde{D}_2 \tilde{D}_3 
+E_3 \tilde{E}_3 \tilde{D}_1 \tilde{D}_2
+D_1' \tilde{D}_1' \tilde{D}_1 \tilde{E}_3  \cr
&+D_2' E_3 \tilde{D}_2' \tilde{E}_3 \tilde{D}_1'
+E_2 \tilde{E}_2\tilde{D}_1' \tilde{D}_3' 
+E_2 D_1 \tilde{E}_2 \tilde{D}_1 \tilde{D}_1' 
+ E_2 D_3 \tilde{E}_2 \tilde{D}_3 \tilde{D}_3' \cr
&+D_1 D_3 E_2  \tilde{D}_1 \tilde{D}_3 \tilde{E}_2 
+D_2 D_2' \tilde{D}_2 \tilde{D}_2' \tilde{E}_1 
+D_2 D_2' E_3  \tilde{D}_2 \tilde{D}_2' \tilde{E}_3 ). \cr}
\eqlabel\fcsqr$$
where
$$\tilde{M}_i =(1-M_i)^{-1} .\eq$$

%============================================================================== 
\newsec{Diophantine inequalities: elementary couplings, relations and Grobner
bases}

%+ needs to change letters ? does not seem necessary

%^Õ Added a ref

We  introduce the idea of the Grobner basis via a simple
example (see also 
 [\ref{R. Froberg, {\it An introduction to Grobner bases}, Wiley, New York
 1997.}]). 
Suppose $R$ is a model for a generating function, where 
$R=Q[x,y,z,t]/I$  and
$I=(xy-t,zy-t)$ is the ideal generated by
$xy-t$ and $zy-t$, with an $\non^2$ grading given by $(1,0), \,(0,1), \,(1,0)$ and 
$(1,1)$ for $x,\, y,\,z$ and $t$. Writing
$\bar x = x+I$ and similarly for the other variables, we have in $R$
that $\bar x \bar y = \bar t$ and $\bar z\bar y =\bar t$. These two expressions
give two {\it re-write rules} : 
%* correction $xy \mapsto z$ -> $xy \mapsto t$:
$xy \mapsto t$
and $zy \mapsto t$. These rules can be used to simplify any 
monomial. The aim is to find a re-write rule which, when iterated, produces
unique representatives for the classes of $I$. If this is the case, then a
vector space basis of $R$ would consist of terms of the form $m+I$ with
$m$ a monomial which is not divisible by any of the left-hand sides of the
rewrite rules. 
 
In the example above, if we had  `good' rewrite rules then a basis for
$R$ would be represented by monomials not containing $xy$ or $zy$, i.e.
monomials of the form either $y^at^b$ or
$x^az^bt^c$. The generating function which counts these monomials
is:
$$
{1\over {(1-AB)}}\left( {B\over{1-B}}  +
{{1}\over{(1-A)^2}}\right),
\eq$$
The exponent of $A$ carries the first grading index and $B$ 
the second. 

However this generating function is not correct. It contains the
term $2A^2B$ corresponding to the monomials  $xt$ and
$zt$. But the polynomial $ z(xy-t) - x(zy-t) = xt-zt$ is also in $I$
and hence in $R$ we have $\bar x\bar t = \bar z\bar t$ and so
the space of grade $(2,1)$ has dimension 1 rather than 2. This
problem can also be seen as a problem with the re-write rules.
If we start with $xyz$ then we can use the first re-write
rule: $xyz \mapsto tz$ or the second: $xyz \mapsto xt$. We 
cannot apply any further re-write rules and so this set of
re-write rules does not produce a unique representative. The
solution is to include the rule $xt \mapsto zt$. This gives
a set of 3 rules: $xy \mapsto t$, $ zy \mapsto t$ and $
xt \mapsto zt$. It turns out that this is a `good' set
 and so a basis for $R$ is given by (the
classes of) monomials of the form $y^at^b$, $x^az^b$ and
$z^at^b$ which gives the generating function:
$$
{1\over {(1-AB)(1-B)}} + {A\over{(1-A)^2}} + {A\over{(1-A)(1-AB)}}
\eq$$

The set of `good' generators, $xy-t,zy-t,xt-zt$ we have found
for $I$ is known as a {\it Grobner basis} [\ref{B. Buchberger, 
Applications of Grobner basis in non-linear computational geometry
in {\it Trends in Computer Algebra, Lecture Notes in Computer Science}
{\bf 296} ed R Jansen (Berlin: Springer) (1989) 52-80.}\refname\Buch].

%%## parag added:
The general procedure for constructing a Grobner basis given a 
set of generating polynomials is as follows. First choose
a {\it term ordering}, which is an ordering on monomials with
the property that any chain $m_1 > m_2 > \dots $ has finite
length. For example we can order the variables
by $x>y>z>t$ and then order all
monomials by the corresponding lexicographic (dictionary) order, for
example:
$x^2y > xyz > y^3 $.
%* I have added z>t.
 For each
generator of our ideal
$I$, select the monomial which is highest with respect to the given
term ordering. This is then the term which appears on the
left of the re-write rule. The lexicographic ordering gives the
first two re-write rules of our example: $xy \mapsto t$ and
$zy \mapsto t$. Next, for each pair of leading terms find the
lowest common multiple and simplify it in the two possible ways.
In this case there is only one pair of leading terms and
the lowest common multiple is $xyz$ which simplifies to
%z made wording clearer
$xt$ and $yt$. Continue to apply the re-write rules 
until the terms  do not simplify further. If the resulting pair of terms
are the same, then proceed to the next pair of leading
terms, otherwise add a new re-write rule. In this case we add 
$xt\mapsto yt$. Proceed until no pair of leading terms gives
a new rule. This is the case for the rules we now have. For example
the two rules $xy\mapsto t$ and $xt \mapsto zt$ appears to give
a new rule by simplifying $ xyt$ to both $t^2$ and $yzt$. 
%+ small rewording:
However the
second term can be further reduced to $ t^2$ and so no new rule
is required. 

%%## minor changes in the first sentence
This  algorithm for computing Grobner bases is known
as Buchberger's [\Buch] algorithm.
Improvements on this basic algorithm mean that it is now feasible to
find Grobner bases for quite large sets of 
generating polynomials. (The web pages of the
computer-algebra information network at the address http://cand.can.nl/CAIN
contain information about many of the programs currently
available.)
%# last sentence in foot.

%((((
Although it is not clear from this example,  Grobner bases are a
very versatile tool for performing explicit calculations. 
We end this section with an illustrative example  relevant to our discussion
of tensor-product generating functions.

Consider a set of linear Diophantine equations:
$$
M\alpha = 0,\qquad \alpha \geq 0
\eq$$
with $M$ an integer matrix and $\alpha$ a vector of non-negative integers.
We would like to construct a generating function for the solutions to
this set of equations:
$$
\sum_{\alpha} x^\alpha.\eq$$
A non-trivial example is given by
 the Diophantine equations that describe
 a $3\times 3$ magic square:
$$\pmatrix { a&b&c\cr d&e&f\cr g&h&i}\eq
$$
with non-negative entries and equal row and column sums. The magic square
condition
%ý add
(the sum of each row and each column is the same, say equal to $t$) gives the
following set of equations:
$$
\eqalignD{
& a+b+c = t\qquad\quad  &a+d+g = t\cr
& d+e+f = t\qquad \quad  &b+e+h = t\cr
& g+h+i = t\qquad \quad  & c+f+i = t\cr}
\eq$$
With $\alpha$ standing for the column vector with entries $(a,b,c,d,e,f,g,h,i,t)$,
the matrix
$M$ reads
$$M= \pmatrix{1&1&1&0&0&0&0&0&0&-1\cr
0&0&0&1&1&1&0&0&0&-1\cr
0&0&0&0&0&0&1&1&1&-1\cr
1&0&0&1&0&0&1&0&0&-1\cr
0&1&0&0&1&0&0&1&0&-1\cr
0&0&1&0&0&1&0&0&1&-1\cr
}\eq$$

There is a straightforward
algorithm for finding the basic set of solutions 
[\Huet] which yields:
$$\eqalignD{
&  \alpha_1 = (0, 0,1,0 ,1,0,1,0,0,1)\qquad \quad &
    \alpha_4 = (1  , 0 ,  0  , 0 ,  0 ,  1 ,  0 ,  1 ,  0 ,  1 )\cr
&   \alpha_2 = (0 ,1, 0 ,  0 ,  0 ,  1 ,  1 ,  0 ,  0  , 1 )\qquad\quad  &
   \alpha_5 = (0  , 1 ,  0 ,  1 ,  0 ,  0 ,  0 ,  0 ,  1 ,  1 )\cr
&   \alpha_3 = (0  , 0 ,  1 ,  1 ,  0 ,  0 ,  0 ,  1 ,  0 ,  1 )\qquad \quad &
  \alpha_6 = (1  , 0 ,  0 ,  0 , 1 ,  0 ,  0 ,  0  , 1  , 1)\cr}
\eq$$

We shall use $A,B,\dots,T$ to denote the ``grading variables''
of this example so that the exponent of $A$ carries
the value of $a$ and so on.
A model for the generating function is given by the
subring $S$ of $\Q[A,B,C,D,E,F,G, H,I,T]$ generated by monomials
corresponding to the 6 elementary solutions, 
$$
\eqalignT{
E_1&=CEGT,\qquad 
 E_2&=BFGT,\qquad 
 E_3&=CDHT,\cr
 E_4&=AFHT,\qquad 
 E_5&=BDIT,\qquad
 E_6&=AEIT\cr}
\eq$$
The monomials in $S$ correspond to magic squares. For example
$E_1^2E_4E_6 = A^2C^2E^3FG^2HIT^4\in S$ corresponds to a square
with row and column sums equal to 4:
$$
\pmatrix{ 2&0&2\cr 0&3&1\cr 2&1&1\cr}.
\eq$$
Note that in this example it is convenient
to construct our model as a subring
 of the ring of grading variables. 
Thus each ``elementary coupling'' $E_i$ is actually
equal to the corresponding monomial in the grading
variables.

However, there are relations between these generators
and so it is not immediately clear how to construct the Poincar\'e series
for $S$. What we require is an isomorphism of $S$ with $R = \Q[e_1,\dots,e_6]/I$
such that $e_i\mapsto E_i$, $i=1,\dots,6$ and such that we have a Grobner
basis of the ideal $I$ (the `ideal of relations').

Fortunately, such an isomorphism is easily constructed using Grobner-basis
methods. Introduce the ring $\Q[A,B,C,D,E,F,G,H,I,T,e_1,\dots,e_6]$
with the lexicographic ordering $$A>B>C>D>E>F>G>H>I>T>e_1>\dots > e_6\eq$$
Let $J$ be the ideal generated by $E_1-e_1,\dots, E_6-e_6$. This is
not necessarily a Grobner basis with respect to this term ordering. Let $G$ be the Grobner basis
for $J$ with the given ordering. Then it can be shown [\Buch] that $G\cap \Q[e_1,\dots,e_6]$
is a Grobner basis for the ideal of relations $I$ which we require. 
In this case $G$ is quite large, but its intersection with $\Q[e_1,\dots,e_6]$
is $e_1e_4e_5-e_2e_3e_6$. The corresponding relation in $R$ is
 $E_1E_4E_5-E_2E_3E_6$
%((((
and  these two
terms do indeed give the same magic square, so that indeed
we have found a relation between the generators of $R$. The Poincar\'e series
for $\Q[e_1,\dots,e_6]/I$ is easily computed:
%$$\eqalign{
%&{1\over{(1-BFGT)(1-CDHT)(1-AEIT)}}\cr &\times 
%\left({1\over{(1-CEGT)(1-AFHT)}}+ {{BDIT}\over{(1-CEGT)(1-BDIT)}} \right.\cr
%&\qquad+ \left.
%{{AFHTBDIT}\over{(1-AFHT)(1-BDIT)}}\right)\cr}
%\eq$$
%
$$ \eqalign{
{1\over(1-E_2)(1-E_3)(1-E_6)} &\left(
{1\over{(1-E_1)(1-E_4)}} \right.\cr  & \, +
\left.{{E_5}\over{(1-E_1)(1-E_5)}} +{{E_4E_5}\over{(1-E_4)(1-E_5)}}\right)\cr}
\eqlabel\majgen
$$

%============================================================================== 

\newsec{Berenstein-Zelevinsky Triangles}

%------------------------------------------------------------------

\subsec{Generalities}

The previous examples make clear the usefulness
of a re-expression of the tensor-product calculation in
terms of Diophantine  inequalities.  The Littlewood-Richardson algorithm yields a
set of such inequalities only for $su(N)$.  Fortunately, 
Berenstein and Zelevinsky [\ref{ A.D. Berenstein and
A.V.  Zelevinsky, J. Geom. Phys. {\bf 5} (1989) 453.}\refname\BZ] have expressed
the solution of the multiplicity of a given tensor product as a counting problem
for the number of integral points in a convex polytope.  For a given algebra, the
polytope is formulated in terms of a characteristic set of inequalities.  
For
$su(N)$, these reduce to the LR set of inequalities.  For the other
classical algebras, except $sp(4)$, the proposed set of inequalities is a conjecture.

%------------------------------------------------------------------

\subsec{BZ triangles for $sp(4)$}

The combinatorial description of tensor products for $sp(4)$ is not as simple as
in the $su(N)$ case:
%z removed next line, not really clear, description of procedure is
%z below. Also modification rules are needed.
%z since the plain application of a modified LR algorithm is
%z not linear:
%" ok  
a standard LR product must be supplemented by a division
operation and modification rules [\ref{G.R.E. Black, R.C  King and B.G.  Wybourne, J. Phys. A: Math.
Gen. {\bf 16} (1983) 1555.}\refname\BGW]. 
Given the BZ set of inequalities, the natural way to proceed, as just mentioned,
 is to interpret these as the appropriate inequalities for the description of the
tensor products. These inequalities are as follows: 
$$
\eqalignD{&\lambda_1  \geq p \qquad &\mu_1  \geq q \cr
&\lambda_2  \geq r_{1}/2   \qquad &\mu_1  \geq q+r_1-r_{2} \cr &\lambda_2  \geq
r_{1}/2+q-p
 \qquad & \mu_1  \geq p+r_1-r_{2} \cr &\lambda_2  \geq r_{2}/2+q-p
\qquad & 
\mu_2  \geq r_{2}/2 \cr
&\nu_1  = r_2-r_1-2p+\la_1+\mu_1\; \qquad &
\nu_2  = p-q-r_2+\la_2+\mu_2\cr}
 \eqlabel\bzspq $$
%%## change the place of insertion of the footnote:
(Our notation is different
from that used in [\BZ]: the relation is $r_1=m_1,\, r_2=m_2,\,
p=m_{12},\,q=m_{12}^\dagger$.)
The $sp(4)$ tensor product coefficient  ${\cal
N}_{\lambda\mu\nu}$ is thus given by the number of solutions of the above system
with
%(((( I don't think we defined Z_+. Should it be \N? or \Z^{>0}?
 $r_1,r_2 \in 2~{\N}$ et $p,q \in {\N}$ ($\N$ being the set of nonnegative
integers).

%%%% Ancienne notation%%%%%
% $$
% \eqalign{\lambda_1 & \geq m_{12} \cr
% \lambda_2 & \geq m_{1}/2 \cr \lambda_2 & \geq m_{1}/2+m_{12}^{\dagger}-m_{12}
% \cr \lambda_2 & \geq m_{2}/2+m_{12}^{\dagger}-m_{12} \cr 
% \mu_1 & \geq m_{12}^{\dagger} \cr
% \mu_1 & \geq m_{12}^{\dagger}+m_1-m_{2} \cr
% \mu_1 & \geq m_{12}+m_1-m_{2} \cr
% \mu_2 & \geq m_{2}/2 \cr
% \nu_1 & = m_2-m_1-2m_{12}+\la_1+\mu_1\cr
% \nu_2 & = m_{12}-m^\dagger_{12}-m_2+\la_2+\mu_2\cr}
%  \eqlabel\bzspq $$
% The $sp(4)$ tensor product coefficient  ${\cal
% N}_{\lambda\mu\nu}$ is thus given by the number of solutions of the above system
% with
%  $m_1,m_2 \in 2~{\bf Z}_{+}$ et $m_{12},m_{12}^{\dagger} \in
% {\bf Z}_{+}$.

%%% No need for the $g_i$:
%g_1= & (\lambda_1+\nu_1-\mu_1)+(\lambda_2+\nu_2-\mu_2)
% =m_{12}+m_{12}^{\dagger}+m_1 \cr
% g_2= & {(\lambda_1+\nu_1-\mu_1) \over 2}+(\lambda_2+\nu_2-\mu_2)=
% m_{12}^{\dagger}+m_1/2+m_2/2 \cr}
%%%%%%

A proper set of variables for a complete description of a particular
tensor-product coupling is thus
$\{\la_1,\la_2,\mu_1,\mu_2, r_1,r_2,p,q\}$. We give the list of
elementary couplings, adding to each coupling the corresponding four-vector $[
r_1,r_2,p,q]$:
$$\eqalignT{
&A_1:  &(0,0)\otimes(1,0)\supset(1,0)\quad [0,0,0,0]\cr
&A_2:  &(1,0)\otimes(0,0)\supset(1,0)\quad [0,0,0,0]\cr
&A_3:  &(1,0)\otimes(1,0)\supset(0,0)\quad [0,0,1,1]\cr
&B_1:  &(0,0)\otimes(0,1)\supset(0,1)\quad [0,0,0,0]\cr
&B_2:  &(0,1)\otimes(0,0)\supset(0,1)\quad [0,0,0,0]\cr
&B_3:  &(0,1)\otimes(0,1)\supset(0,0)\quad [2,2,0,0]\cr
&C_1:  &(0,1)\otimes(1,0)\supset(1,0)\quad [0,0,0,1]\cr
&C_2:  &(1,0)\otimes(0,1)\supset(1,0)\quad [0,2,1,0]\cr
&C_3:  &(1,0)\otimes(1,0)\supset(0,1)\quad [0,0,1,0]\cr
&D_1:  &(2,0)\otimes(0,1)\supset(0,1)\quad [0,2,2,0]\cr
&D_2:  &(0,1)\otimes(2,0)\supset(0,1)\quad [2,0,0,0]\cr
&D_3:  &(0,1)\otimes(0,1)\supset(2,0)\quad [0,2,0,0]\cr}\eqlabel\spele$$

%%% Ancienne notation pour le para precedent:
% A proper set of variables for a complete description of a particular
% tensor-product coupling is thus
% $\{\la_1,\la_2,\mu_1,\mu_2,m_1,m_2,m_{12},m^\dagger_{12}\}$.
% We give the list of
% elementary coupling, adding to each coupling the corresponding four-vector $[
% m_1,m_2,m_{12},m^\dagger_{12}]$:

The unspecified linear relations mentioned in (\sprel) can now be
obtained.   To find those products that are  equal in 
the current situation 
we need only compare their corresponding sets of four-vectors
$[r_1,r_2,p,q]$ (which are additive in products of
couplings).  We thus find for instance that 
$$C_1C_2= A_3D_3:\; [0,2,1,1]  \qquad \not= A_1A_2B_3 :\; [2,2,0,0]\eq$$
Proceeding in this way for the other cases, we find the following  complete
list of relations:
$$\eqalignT{ &C_1 C_2 = A_3 D_3,\quad &C_2C_3= A_1D_1\quad & C_3C_1= A_1A_3B_2
\cr &D_1 D_2=B_3C_3^2\quad & D_2D_3= A_1^2B_2B_3\quad & D_1D_3= B_2C_2^2 \cr
&C_1D_1= A_3B_2C_2\quad &C_2D_2= A_1B_3C_3\quad & C_3D_3= A_1B_2C_2\cr}
\eqlabel\zysp$$

The use of the BZ inequalities to find the elementary couplings 
and their relations
%%## change "seems to be" by "is"
 is novel. 
%%## change the next sentence:
(An off-shoot of our construction is that it provides an indirect proof of the
validity of the BZ inequalities since  we recover from it the result of [\Hongo]
derived from the character method.)

A possible choice of forbidden products is the one given in [\Hongo]: $$\{ C_i
C_j,D_i D_j,C_i D_i\}\eq$$ with $i,j=1,2,3~ {\rm and} ~i\not=j$.
It leads to the generating function:
$$\eqalign{G^{sp(4)}= \left(\prod_{i=1}^3~\tilde{A}_i
\tilde{B}_i\right) &\left( \tilde{C}_1 \tilde{D}_2+D_3 \tilde{C}_1
\tilde{D}_3+C_2 D_1 \tilde{C}_2 \tilde{D}_1 \right.\cr &\left. +C_2 \tilde{C}_2
\tilde{D}_3+D_1
\tilde{C}_3 \tilde{D}_1+C_3 \tilde{C}_3 \tilde{D}_2\right) \cr} \eqlabel\fcr$$
Of course, by modifying the ordering in the Grobner basis, we can get other
choices of forbidden couplings.  Here is another set of forbidden couplings that
can be obtained: $\{D_i D_j, C_i D_i,A_1 D_1,A_3 D_3, A_1 A_3 B_2\}$.
The corresponding generating function reads
$$\eqalign{G^{sp(4)}&=\tilde{B}_1 \tilde{B}_2 \tilde{B}_3
 \left[\left(\prod_{i=1}^3~\tilde{A}_i\right) \tilde{C}_i (1-A_1 A_3 B_2)+D_3
\tilde{D}_3 \tilde{A}_1 \tilde{A}_2 \tilde{C}_1 \tilde{C}_2 \right.\cr
&\quad\left.+D_1 \tilde{D}_1 \tilde{A}_2 \tilde{A}_3 \tilde{C}_2 \tilde{C}_3 
 + D_2 \tilde{D}_2 \tilde{A}_1 \tilde{A}_2 \tilde{A}_3 \tilde{C}_1
\tilde{C}_3 (1-A_1 A_3 B_2) \right] .\cr}
\eq$$
These two generating functions are equivalent when rewritten in terms of the
grading variables, that is, in terms of Dynkin labels.  However, they originate
from two distinct models.  The second one turns out to be well adapted to the
fusion extension.

%================================================================================

%# modifications here: intro paragraph

\newsec{A vector basis approach to the construction of generating
functions}

%##added the Stanley ref here
%ý OK

%ýý now we take out the ref to stanley in the text and push it to a footnote

In this section, we present a simple and systematic way
%, described  in the first reference in [\Stan],
 of generating by hand
all the elementary solutions of a set of linear homogeneous inequalities starting
from the well-known construction of a vector basis.  The first step
amounts to reformulate the system of inequalities in terms of equalities.  We then
look for the elementary independent solutions by relaxing the positivity
requirement.  In  other words, we construct the vector basis.  In a final step,
we find the minimal linear combinations of these vector basis elements that yield
positive solutions. This will also provide an illustration of
MacMahon's projection technique.  The result of this projection is
the desired tensor-product generating function.  Hence, this approach turns out
to be a new way of constructing the tensor-product generating functions. 
%ýý the foot is introduced to clarify the 
%ýý "novel" status of our approach
%*% reformulation of the footnote
(This generic method, referred to as being novel for tensor products,
 is certainly well-known in general: 
 it is  discussed in the first reference of [\Stan].)
%ýý add:

%------------------------------------------------------------------------
\subsec{Graphical
 representations as BZ triangles for
 $su(N)$}

%\subsec{From the LR rule to BZ triangles}

Consider the direct transformation of
the LR inequalities to equalities by introducing an appropriate number of new
%z changed positive-integer to non-negative, Is this right?
%" yes!
non-negative integer variables.  Consider first the $su(2)$ case, for
which there is a single inequality: $\la_1\geq n_{12}$.  We transform this into
%(((( changed positive to non-negative
an equality by introducing the non-negative integer $a$ defined by
$$\la_1= n_{12}+a \eq$$ The expression for $\nu_1$ becomes then
$\nu_1=\la_1+n_{11}-n_{12}=a+n_{11}$.  Since $\mu_1=n_{11}+n_{12}$, we are led
naturally to a triangle representation of the tensor product:
$$\la\otimes \mu\supset \nu \quad \leftrightarrow \quad \matrix{a \cr
n_{12} \quad n_{11} \cr} \quad \eqlabel\wsd$$
We read off the Dynkin label of the $\la$ representation from the sum of the two
integers that form the left side of the triangle, that of the
$\mu$ representation from the bottom of the triangle and the
$\nu_1$ label is the sum of the two integers that form the right side.  A more
%# rewording:
uniform notation amounts to setting $a= m_{12}$ and $n_{11}= l_{12}$, in
terms of which the triangle looks quite symmetrical:
$$\matrix{m_{12} \cr
n_{12} \quad\quad l_{12} \cr} \quad \eqlabel\wsdd$$
 with
$$\la_1= m_{12}+n_{12}\qquad \mu_1= n_{12}+l_{12}\qquad \nu_1=m_{12}+l_{12}\eq$$
These numbers $m_{12}$ and $l_{12}$ plays the role of $n_{12}$ in the permuted
versions of the tensor product.
The triangle combinatorial reformulation of the tensor product problem is as
follows: the number of triangles that can be formed from non-negative
integers $n_{12}, \,m_{12}$ and $l_{12}$ that add up to the Dynkin
labels of the representations under study according to the above relations gives
the multiplicity of the triple coupling $\la\otimes \mu\supset \nu$, or
equivalently, the multiplicity of the scalar representation in the product
$\la\otimes \mu\otimes \nu\supset (0)$ (since for $su(2)$, $\nu^*=\nu$).

 The situation for $su(3)$ is somewhat more complicated.  The transformation
of the LR inequalities (\nijrang, \nijlr) into equalities in this case takes the form 
$$
\eqalignD{ 
& \la_1  = n_{12}+a\qquad &  n_{11}= n_{22}+d \cr 
&  \la_2 = n_{13}+b\qquad & n_{11}+n_{12}= n_{22}+n_{23}+e\cr
& \la_2+n_{12} =n_{13}+n_{23}+c\qquad &~\cr}
\eqlabel\eqee$$
The expression for the other weights becomes
$$\eqalignD{ &\mu_1 = n_{13}+e\qquad
&\mu_2 = n_{22}+n_{23}\cr
&\nu_1 = a+d\qquad
&\nu_2 = n_{22}+c\cr}\eqlabel\others$$
Since there are two expressions for both $n_{11}$ and $\la_2$,
there follows the compatibility relations:
$$n_{12}+d= n_{23}+e\qquad n_{23}+c= b+n_{12}\eq$$
By adding these two  relations, we find:
$$c+d=b+e\eq$$
Again we are led naturally to a triangle representation: with $\zeta= \nu^*$ this
reads
$$\matrix{a_{~}\cr
	n_{12}~~\quad d_{~}\cr
b_{~}~\quad\qquad ~~c_{~}\cr
 n_{13}~\quad e \qquad n_{23} \quad~ n_{22} \cr
}\eqlabel\dtribzz$$
We read the Dynkin labels from the sides of the triangles, from $\la_1$ to
$\zeta_2$ in an anti-clockwise rotation starting from the top of the triangle,
exactly  as for
$su(2)$, except that here there are two labels on each sides.  Notice
that the compatibility conditions amounts to the equality of the sums of
the extremal points of the three pairs of opposite sides of the hexagon obtained
by dropping the three corners of the triangle.

 Again a more symmetrical notation
is:
$$a= m_{13}\qquad b= m_{23}\qquad  c= m_{12}\qquad  d= l_{23}\qquad  e= l_{12}\qquad 
 n_{22}= l_{13}\qquad\eq$$
in terms of which the triangle reads
$$\matrix{m_{13}\cr
	n_{12}~~\quad l_{23}\cr
m_{23}~\quad\qquad ~~m_{12}\cr
 n_{13}~\quad l_{12} \qquad n_{23} \quad~ l_{13} \cr
}\eqlabel\trbz$$ 
with labels fixed by: 
$$\eqalignD{&\la_1=m_{13}+n_{12}\quad &\la_2= m_{23}+n_{13}\cr
&\mu_1= n_{13}+l_{12} \quad &\mu_2= n_{23}+l_{13}\cr
&\zeta_1= l_{13}+m_{12}\quad &\zeta_2= l_{23}+m_{13}\cr}
\eqlabel\wee$$
The hexagon conditions read:
$$\eqalign{n_{12}+m_{23}& = n_{23}+m_{12},\cr
l_{12}+m_{23} &= l_{23}+m_{12},\cr 
l_{12}+n_{23} &= l_{23}+n_{12}.\cr
}\eqlabel\dtribz$$
In terms of triangles, the problem of finding the multiplicity of the $su(3)$
tensor product $\la\otimes \mu\otimes \zeta\supset 0$ boils down to enumerating
the number of triangles made with non-negative integers that form a bipartition of
the Dynkin labels and that  satisfy the above three  hexagon relations. (For the
$su(N)$ generalisation, see [\ref{L. B\'egin,
A.N. Kirillov, P. Mathieu and M. Walton, Lett. Math. Phys. {\bf 28} (1993)
257.}\refname\BKMW]).
%%## add a ref above 

 Here is the rationale for
the labelling $n_{ij},m_{ij},l_{ij}$ from the triangle point of view [\ref{A.D. Berenstein and A.Z.
Zelevinsky, J. Algebraic Combinat. 1 (1992) 7.}\refname\BZtri]. If
$e_i$ are orthonormal vectors in ${\bf R}^N,$ then the positive roots of $su(N)$
can be represented in the form $e_i-e_j,\ 1\leq i<j\leq N.$ The triangle encodes
three sums of positive roots:
$$\eqalign{\mu +\zeta -\lambda^*\ &=\ \sum_{i<j}\ l_{ij} (e_i-e_j)\ \ ,\cr
\zeta +\lambda -\mu^*\ &=\ \sum_{i<j}\ m_{ij} (e_i-e_j)\ \ ,\cr
\lambda +\mu -\zeta^*\ &=\ \sum_{i<j}\ n_{ij} (e_i-e_j)\ \ ,\cr}
\eq$$
 The hexagon relations are simply the consistency
conditions for these three expansions. Clearly, the variables $n_{ij}$ that
appear in the above relations are exactly the $n_{ij}$ that appear in the LR
tableaux for the product $\la\otimes \mu\supset\zeta^*=\nu$.

%========================================================================

\subsec{From a vector basis to the generating function: the $su(3)$ case}

%In this section, we present a new method for constructing the tensor-product
%generating functions (which turns out to be described in the first reference in
%##I don't think we can describe the method as "new" if
%##it has already been described by Stanley!
%ý Agree

%ýý wording is changed:
%ýý a foot has previously been introduced to clarify the 
%ýý "novel" status of our approach

Given the transcription of inequalities into equalities, we can easily extract
the corresponding basis vectors.  This is the starting point of a new method
for constructing the tensor-product generating functions.
%ý cut: (which is described in the first reference in [\Stan])
To keep things
concrete, we focus on the $su(3)$ case.  The goal is to first get a vector basis
and then to project it to get the elementary couplings.  The generating function
is a direct result  of this procedure.

The equality version of 
the LR inequalities 
are (\wee) and (\dtribz); they underlie the construction of the BZ triangle
(\trbz). 
The last hexagon condition of (\dtribz) is the difference of the previous two so
it is not an independent relations. We thus have a total of 15 variables: $\la_1,
\cdots,
\zeta_2, l_{12}, \cdots , n_{23}$ and 8 equations. The number of independent
variables is thus 7. These will be chosen to be
$m_{13},m_{23},l_{13},l_{23},n_{12},n_{13},n_{23}$. The dependent variables are
fixed as follows:
$$\eqalignD{ 
& \lambda_1= m_{13}+n_{12} \qquad 
 & \lambda_2= m_{23}+n_{13}  \cr
 &  \mu_1 = n_{13}+n_{12}+l_{23}-n_{23}			\qquad 
 &  \mu_2 = n_{23}+l_{13}				\cr
 &  \zeta_1 = n_{12}+m_{23}+l_{13}-n_{23}			\qquad 
	& 	\zeta_2 = l_{23}+m_{13}				\cr
 &  l_{12} = n_{12}+l_{23}-n_{23}				\qquad 
 &  m_{12} = n_{12}+m_{23}-n_{23}				\cr} \eq $$
We now look for the elementary solutions of this system (without invoking
the constraint that all the above dependent variables should be necessarily
positive).  The sought basis vectors are obtained by setting one of the
variable 
$m_{13},\cdots,n_{23}$ to 1 and all other set equal to zero.
%*% add a footnote

 This produces (in
order) the triangles
$E_2, E_5, E_6, E_3, E_7, E_4$ and
$Z_1$ displayed below:
$$\matrix{E_2: (1,0)(0,0)(0,1)\cr~\cr\tri000100000}\qquad
\matrix{E_3: (0,0)(1,0)(0,1)\cr~\cr\tri000010001}$$
$$\matrix{E_4: (0,1)(1,0)(0,0)\cr~\cr\tri100000000}\qquad
\matrix{E_5: (0,1)(0,0)(1,0)\cr~\cr\tri010001000}\qquad
\matrix{E_6: (0,0)(0,1)(1,0)\cr~\cr\tri000000100}$$
$$\matrix{E_7: (1,0)(1,0)(1,0)\cr~\cr\tri001001001}\qquad
\matrix{Z_1: (0,0)(-1,1)(-1,0)\cr~\cr\tri00000{-1}01{-1} } \eqlabel\zz$$
These are all genuine BZ triangles except for $Z_1$ which has some negative
entries.  However, at this level, there are no relations
between these elementary solutions (the basis vectors are independent), hence
the decomposition of any solution in terms of these 7 basic ones is unique.  All
solutions are then freely generated from the following function:
$$G={1 \over (1-E_2) (1-E_3) (1-E_4) (1-E_5) (1-E_6) (1-E_7) (1-Z_1)}\eq$$
To recover the generating function for all tensor products from the above
expression, we need to project out terms that lead to triangles with negative
entries.  To achieve this, we introduce the grading variables associated to the
above couplings (compare the above triangles with the general form given in
(\trbz)):
$$\eqalignT{ &E_2: M_{13}, ~\qquad &E_3: L_{12}L_{23}\qquad &E_4: N_{13}\cr
 &E_5: M_{12}M_{23}\qquad &E_6: L_{13}\qquad &E_7: L_{12}M_{12}N_{12}\cr
& ~ &Z_1:L_{12}^{-1}M_{12}^{-1}N_{23}  &\cr}\eq$$
Our generating function follows from the projection of the above function $G$, 
re-expressed in terms of the grading variables, to positive powers of $L_{12}$ and
$M_{12}$.  Equivalently, one can re-scale $L_{12}$ by $x$ and $M_{12}$ by $y$ and
project to positive powers of $x$ and $y$ and set $x=y=1$ in the result.  This is
equivalent to the rescaling
$$E_3\rw xE_3\qquad E_5\rw yE_5\qquad E_7\rw xyE_7\qquad Z_1\rw
x^{-1}y^{-1}Z_1\eq$$
We are thus led to consider 
$$\Ox\Oy \; G(E_2, xE_3, \cdots, x^{-1}y^{-1}Z_1)\eq$$
Keeping only those terms which depend explicitly upon $x$ or $y$, we have then
$$\eqalign{\Ox\Oy \; &{1 \over  (1-xE_3)(1-yE_5)  (1-xyE_7)
(1-x^{-1}y^{-1}Z_1)}\cr
& \quad= {1 \over  (1-xE_3)(1-yE_5)  (1-E_7Z_1)}\left({1\over
1-xyE_7}+{x^{-1}y^{-1}Z_1\over 1-x^{-1}y^{-1}Z_1}\right)\cr}\eq$$
No more work is needed for the first term. For the second one, we have
$$\eqalign{\Ox\Oy \; &{x^{-1}y^{-1}Z_1 \over  (1-xE_3) (1-E_7Z_1)(1-x^{-1}Z_1E_5)
}\left({yE_5\over 1-yE_5}+{1\over 1-x^{-1}y^{-1}Z_1}\right)\cr
&=\Ox\; {x^{-1}E_5Z_1 \over (1-E_5)(1-E_7Z_1) (1-xE_3) (1-x^{-1}Z_1E_5)
}\cr
&=\Ox\; {x^{-1}E_5Z_1 \over (1-E_5)(1-E_7Z_1) (1-E_3E_5Z_1)}\left({xE_3\over 
1-xE_3}+{1\over 1-x^{-1}Z_1E_5 }\right)\cr
&= {E_3E_5Z_1\over (1-E_5)(1-E_7Z_1) (1-E_3E_5Z_1)(1-E_3)}\cr}\eq$$
We then introduce  the following two new elementary couplings
$$E_1=E_7 Z_1\; ,   \qquad \quad E_8= E_3 E_5 Z_1\eq$$
Collecting the two terms resulting from the projection, we end up with
$$G^{su(3)}= \left(\prod_{i=1}^8 \tilde{E}_i\right) (1-E_7E_8)\eq$$
which is indeed the $su(3)$ tensor-product generating function.

It is worth
pointing out that the Elliot-MacMahon algorithm that has been presented here
as a method distinct from the vector basis, can be
reinterpreted in a way that makes the two approaches equivalent.  This is done
in  section 3 of the first reference of [\Stan]. There, the elementary
solutions are not obtained as
 above by setting successively one dependent variables equal to 1 and
the others equal to 0, but in reading them off directly from the columns of the
$8\times 7$ matrix of the matrix version of the above equation:
$$
\pmatrix{
1&1&0&0&0&0&0\cr
0&0&1&1&0&0&0\cr
0&1&0&1&-1&0&1\cr
0&0&0&0&1&1&0\cr
0&1&1&0&-1&1&0\cr
1&0&0&0&0&0&1\cr
0&1&0&0&-1&0&1\cr
0&1&1&0&-1&0&0\cr} \pmatrix
{m_{13}\cr n_{12}\cr m_{23}\cr n_{13}\cr n_{23}\cr l_{13}\cr l_{23}\cr}= 
\pmatrix
{\la_{1}\cr \la_{2}\cr \mu_{1}\cr \mu_{2}\cr \zeta_{1}\cr \zeta_{2}\cr l_{12}\cr
m_{12}\cr}$$
The exponentiated version of the columns gives the elementary solutions written
below.  This leads to the so-called `crude' generating function that is then
projected onto the positive solutions by the usual method.

%==============================================================================
\subsec{General aspects of the vector basis construction} 

%# this section has been modified a bit

In general, of
course, the fundamental solutions to the linear system may have non-integral
values of the variables. However the corresponding terms
in the generating function can be eliminated by
rationalising all the denominator terms and then
keeping only those terms in the numerator that
have integral exponents. This suggests the following
modification of MacMahon's algorithm.

%## I haven't look at the Stanley ref yet. Is this exactly what
%## he describes? If so we should say so. 
%ý No this is not in Stanley (he simply gives the idea
%ý of the vector basis: so keep it in this form.

Consider the system of equations
$$Mx=0,\quad x\in\non^k\eqlabel\Meq$$
where $M$ is a matrix of rank $s$.  We thus have $k$ variables and $s$ relations
between them. The dimension of the vector basis is thus $k-s$.  We will denote
the independent (free) variables as $x_i$, $i=1, \cdots, k-s$ and the remaining
ones as ${\tilde x}_j$, $j=1, \cdots, s$.  To find a generating function for the
solutions of this system:

%z Changed working to show we have to construct the basis in a particular way
\item{\bf 1.} First construct a basis in $\Q^k$ for the solutions
of $Mx=0$ 
% Using the usual linear algebra method
% we can arrange for the solutions in the basis to
% have the form:
%# more explicit
by setting $x_i=1$ with all other $x_j$ zero ($j=1, \cdots, k-s,
\, j\not=i$).  Denote by
${\tilde x}_j^{(1)}$ the value of the dependent variable ${\tilde x}_j$
evaluated at
$x_1=1$ with all other $x_i$ zero. The basis then reads
%# change e_i for \epsilon_i since e_i are used for ele.coupl.
$$
\eqalign{ &\epsilon_1=(1,0,0\dots, 0; \{{\tilde x}_j^{(1)}\}),\cr
 &\epsilon_2=(0,1,0\dots, 0; \{{\tilde x}_j^{(2)}\}),\cr
 &\cdots\cr
&\epsilon_{k-s}=(0,0,0,\dots,1; \{{\tilde x}_j^{(k-s)}\})\cr}\eq
$$
By construction, the $\epsilon_i$'s are linearly independent.  
However notice that in general the ${\tilde x}_j^{(i)}$ might be rational.
%z next phrase is redundant
%" ok
%z that is ${\tilde x}_j^{(i)} \in \QQ$.

\item{\bf 2.} From the form of the $\epsilon_i$'s, it follows that
any solution to (\Meq) can be written as $\sum_i c_i\epsilon_i$
with $c_i$ non-negative integers. In particular this means
that every solution to (\Meq) corresponds to
a term in the generating function:
$$G(X)={1\over{(1-X^{\epsilon_1})(1-X^{\epsilon_2})\dots(1-X^{\epsilon_s})}}\eq$$
where $X_1,\dots,X_k$ are grading variables.

\item{\bf 3.} $G(x)$ may contain negative or fractional
exponents due to the occurrence of ${\tilde x}_j^{(i)}$ in the exponents. These are
eliminated by first using MacMahon's algorithm to eliminate any negative
exponents and then rationalising denominators and keeping only terms
with integral exponents in the numerators.

The result is the generating function for the solutions
to (\Meq). This algorithm, however, does not seem to be
optimal in all case.
%==================================================================

\subsec{Multiple $su(2)$ products from the vector basis construction}

A simple and different application of the formalism just developed is furnished
by the analysis of $su(2)$ quadruple tensor products.  This application is
different in that it does not rely on the triangle description and as such, its
formulation is less direct.
%ýý add a foot
%(((( removed ( )
This does not mean however that there are no diagrammatic representations
for the quadruple product. 
In fact, having a set of inequalities, we can transform then into equalities, as
it is done below, and from them set up a diagrammatic representation. In the
present case, it would correspond to two adjacent
$su(2)$ triangles, one upside down, with their adjacent sides forced to be
%((((
equal.  However, our analysis will not rely on such
a description.
 It will serve as a preparation the somewhat more complicated
$sp(4)$ example treated in the following section. 

The Diophantine description of this problem has been presented in section 4.2. It
is based on the two inequalities (\multiin) which are readily transformed into
%((((changed positive to non-negative
equalities by the introduction of two non-negative integers $a_1, \, a_2$:
$$\la_1= n_{12}+a_1\qquad \la_1+n_{11}-n_{12}= m_{12}+a_2\eq$$
However this system does not contain any reference to the variable $m_{11}$ and
for this reason we introduce the further constraint $m_{11}\geq 0$ which calls
%(((( changed changed integer to non-negative integer
for a new non-negative integer variable:
$$m_{11}= a_3\eq$$
We have thus a total of 8 variables : $\{\la_1, 
n_{11}, n_{12},m_{11},m_{12}, a_1, a_2, a_3\}$ and 3 equations.  There are thus 5
independent variables, chosen to be  $\{a_1, a_2, a_3, n_{12}, m_{12}\}$.  The
basis vectors, with components ordered as follows $$(a_1, a_2,
a_3, n_{12}, m_{12}; \la_1, n_{11}, m_{11})\eq$$ are obtained by successively
setting equal to 1 one of $\{a_1, a_2, a_3, n_{12}, m_{12}\}$ and the others
equal to 0.  These basis vectors together with their exponentiated version written
in terms of appropriate grading variables read:
$$\eqalignT{
& (1,0,0,0,0;1,-1,0)\quad &: L_1N_{11}^{-1}\Ac_1\cr
& (0,1,0,0,0;0,1,0)\quad &: N_{11}\Ac_2\cr
& (0,0,1,0,0;0,0,1)\quad &: M_{11}\Ac_3\cr
& (0,0,0,1,0;1,0,0)\quad &: L_1N_{12}\cr
& (0,0,0,0,1;0,1,0)\quad &: N_{11}M_{12}\cr
}\eq$$
The desired generating function is obtained from the projection to positive powers
of
$N_{11}$ of the function
$${1\over (1-L_1N_{11}^{-1}\Ac_1)(1- N_{11}\Ac_2)(1-L_1N_{12}) (1-
N_{11}M_{12}) (1- M_{11}\Ac_3) }\eq$$
The projection operation is done by the familiar method and the result, after
setting all $\Ac_i=1$ is
$$ G= {1-L_1N_{11}M_{12}\over  (1- L_1N_{12})(1-
L_1M_{12})(1-L_1) (1-N_{11}M_{12})(1- N_{11} )(1- M_{11})}\eq$$ from which we
read of the 6 elementary couplings $E_1, \cdots , E_6$ (in the order where they
appear in the denominator) given in (\mutipo) and the relation
$E_3E_4= E_2E_5$.  The above function is exactly the one derived in section 4.2.

%# Chris: this calculation has been done.

%------------------------------------------------------------------------

\subsec{$sp(4)$
diamonds and the vector basis derivation of the generating function}

The system of inequalities (\bzspq ) pertaining to $sp(4)$ can be transformed
into a system of equations in the standard way: by setting 
$r_1/2=s_1$ and $r_2/2=s_2$ and 
%(((( changed integer to non-negative integer
introducing the non-negative integers $a_i$, we get
%ý add a foot to thanks mark for old trials on this
%%## put the ref to M Walton in references:
[\ref{The original idea of looking for a
diagrammatic representation of
$sp(4)$ tensor products along theses lines is due to M. Walton.}]: 
$$\eqalignD{
&\lambda_1  =p+a_1 \qquad &\nu_2  =a_4+a_8 \cr
&\lambda_2  =s_1+a_2 \qquad &a_2+p =a_3+q \cr
&\mu_1  =q+a_5  \qquad &	a_3+s_1  =a_4+s_2 \cr
&\mu_2  =s_2+a_8 \qquad &a_5+2s_{2}  =a_6+2s_{1} \cr
&\nu_1  =a_1+a_7 \qquad & a_6+q  =a_7+p \cr}   \eq$$
This leads to a diamond-type graphical representation of the tensor product that
has the advantage over the one presented in [\BZ] of being linear in that the sum
of two diamonds is also a diamond.  This is illustrated in Fig. 1.
%%## minor reformulation of the end of last sentence

In Fig. 1, all data pertaining to the first (second) Dynkin label appear at
the left (right).
Dotted lines relate those two points that compose the label indicated beside it. 
Opposite continuous lines are constrained to be equal, with the length of a line
being defined as the sum of its extremal points except for the lines delimited by
the points
$(a_6, s_1)$ and $(a_5,s_2)$ where the point $s_i$ is
counted twice (the little bar besides $s_1$ and $s_2$ being a
reminder of this).  Explicitly, for those lines, we have thus the
constraint
$a_6+2s_1=a_5+2s_2$.  Given a triple
$sp(4)$ product,  the number of such diamonds that can be drawn with non-negative
entries yields the multiplicity of the product.  
For instance, the two diamonds that describe the triple coupling $(1,1)\otimes
(1,1) \otimes (2,0)$ are shown in Fig 2.

The dimension of the vector basis is 8 (18 variables and 10 equations, the
last four equations above being linearly independent).  As our free variables we
choose the set $\{s_1, s_2, p, q, a_1, a_3, a_6, a_8\}$.  The 8
basis vectors in terms of grading variables are:
$$\eqalignD{
& \Ec_1: L_2M_1^2N_2\Ac_4\Ac_5^2S_1\qquad &\Ec_2:
M_1^{-2}M_2N_2^{-1}\Ac_4^{-1}\Ac_5^{-2}S_2\cr & \Ec_3:
L_1L_2^{-1}N_1^{-1}\Ac_2^{-1}\Ac_7^{-1}P\qquad &\Ec_4: L_2M_1N_1\Ac_2\Ac_7Q\cr
&\Ec_5: L_1N_1\Ac_1\qquad &\Ec_6: L_2N_2\Ac_2\Ac_3\Ac_4\cr
& \Ec_7: M_1N_1\Ac_5\Ac_6\Ac_7\qquad &\Ec_8 : M_2N_2\Ac_8\cr}\eq$$ 
The generating function is obtained by first projecting of the function
$\prod(1-\Ec_i)^{-1}$ to positive powers for each grading variables and then by
setting all grading variables equal to 1 except for $L_i, M_i, N_i$'s.  The
$sp(4)$  elementary couplings are simple products of the
$\Ec_i$'s (the following
$A_{1,2,3}$ should not be confused with the above grading variables):
$$\eqalignT{
&A_1=\Ec_7\qquad &A_2= \Ec_5\qquad & A_3= \Ec_3\Ec_4 \cr 
&B_1=\Ec_8\qquad &B_2= \Ec_6\qquad & B_3= \Ec_1\Ec_2 \cr
&C_1=\Ec_4\qquad &C_2= \Ec_2\Ec_3\Ec_6\Ec_7^2\qquad & C_3= \Ec_1\Ec_3\Ec_7 \cr
&D_1=\Ec_2\Ec_3^2\Ec_6^2\Ec_7^2\qquad &D_2= \Ec_1\qquad & D_3= \Ec_2\Ec_6\Ec_7^2
\cr}\eq$$
The complete list of $sp(4)$ elementary couplings (\spele) are thus recovered.

\newsec{Conclusion}

As was stressed in the introduction, the main purpose of this
work is to prepare the ground for the analysis of fusion rules, which is the
subject of a sequel paper.  In this paper, we have reviewed the existing
techniques for computing tensor-product generating functions and presented a
comparative  assessment of their  virtues and limitations. 
We also  focused on a model formulation linking 
generating functions to Poincar\'e series, an idea  first 
introduced in [\Stan] and  extended in [\CCS].  Our contribution
 has been to rephrase this program more explicitly, clarify some issues 
and to exemplify the procedure with many examples, some of which are new.
%%## add a sentence:
 An
extended version of this article is available on the Los Alamos server [\ref{L.
B\'egin, C. Cummins  and P. Mathieu {\it Generating functions for
tensor products},  hep-th/9811113.}].

\noindent{\bf Acknowledgement:}
We thank R.T Sharp, J. Patera and M. Walton for useful discussions.  L. B.
thanks S. Lantagne and H. Roussel for computing guidance.  
\vskip 0.15cm
\centerline{\bf REFERENCES}
\immediate\closeout\refs \vskip 0.25cm
  \message{References}\input references
\vfill\eject

\input pictex

\def\bpc{\beginpicture}

\def\epc{\endpicture}

\def\figureF{
\bpc
	\setcoordinatesystem units <0.6mm,0.6mm>
	\setplotarea x from -55 to 55, y from -55 to 55
	\setlinear
	\plot 0 55 -55 0 0 -55 55 0 0 55 /
	\plot 5 -25 32.5 2.5  0 35 /
\plot -5 25 -32.5 -2.5  0 -35 /
\plot 32.5 4.5 30 7 /
	\plot 55 2 52.5 4.5 /
	\multiput {$\bullet$} at 0 55 0 -55 55 0 -55 0 0 35 0 -35 32.5 2.5 -32.5 -2.5 5 -25 -5 25 5 2.5 -5 -2.5
/
	\put {$s_1$} [l] at 	57 0
	\put {$s_2$} [l] at 34.5 2.5
	\put {$q$} [r] at -57 0
	\put {$p$} [r] at -34.5 -2.5
	\put {$\lambda_1$} [b] at -18.75 -1.5
	\put {$\mu_2$} [b] at 18.75 3.5
	\put {$\nu_1$} [l] at -3.5 11.25
	\put {$\nu_2$} [r] at 3.5 -11.25
	\put {$\mu_1$} [b] at -25 16
	\put {$\lambda_2$} [t] at 25 -16
	\put {$a_6$} [b] at 0 57
	\put {$a_5$} [b] at 0 37
	\put {$a_2$} [t] at 0 -37
	\put {$a_3$} [t] at 0 -57
	\put {$a_4$} [r] at 3.5 -25
	\put {$a_7$} [l] at -3.5 25
	\put {$a_8$} [b] at 5 4.5
	\put {$a_1$} [t] at -5 -4.5
	\setdashes
	\plot 32.5 2.5 5 2.5 5 -25 /
	\plot -32.5 -2.5 -5 -2.5 -5 25 /
	\setquadratic
	\plot 0 -35 5 -30.5 15 -22 20 -18 25 -14 30 -10.5 35 -7.5 40 -5 45 -2.5 50 -0.5 55 0
/
	\plot 0 35 -5 30.5 -15 22 -20 18 -25 14 -30 10.5 -35 7.5 -40 5 -45 2.5 -50 0.5 -55 0
/
\epc}

%
%ý Chris: if you do NOT want to see the figure, comment the following line
%ý ( this requires large Tex memory )
%\midinsert
\vskip1cm \centerline{ \figureF} \vskip1cm
%\endinsert

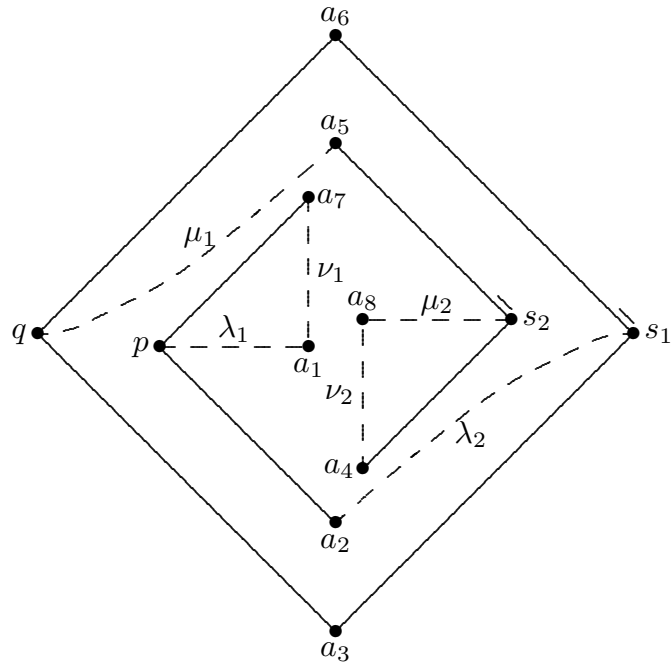
\centerline{Figure 1}
\vfill\eject

\def\figureFpex{
\bpc
	\setcoordinatesystem units <0.6mm,0.6mm>
	\setplotarea x from -55 to 55, y from -55 to 55
	\setlinear
	\plot 0 55 -55 0 0 -55 55 0 0 55 /
	\plot 5 -25 32.5 2.5  0 35 /
	\plot -5 25 -32.5 -2.5  0 -35 /
	\plot 32.5 4.5 30 7 /
	\plot 55 2 52.5 4.5 /
	\multiput {$\bullet$} at 0 55 0 -55 55 0 -55 0 0 35 0 -35 32.5 2.5 -32.5 -2.5 5 -25 -5 25 5 2.5 -5 -2.5
/
	\put {1} [l] at 	57 0
	\put {1} [l] at 34.5 2.5
	\put {0} [r] at -57 0
	\put {0} [r] at -34.5 -2.5
	\put {$\lambda_1$} [b] at -18.75 -1.5
	\put {$\mu_2$} [b] at 18.75 3.5
	\put {$\nu_1$} [l] at -3.5 11.25
	\put {$\nu_2$} [r] at 3.5 -11.25
	\put {$\mu_1$} [b] at -25 16
	\put {$\lambda_2$} [t] at 25 -16
	\put {1} [b] at 0 57
	\put {1} [b] at 0 37
	\put {0} [t] at 0 -37
	\put {0} [t] at 0 -57
	\put {0} [r] at 3.5 -25
	\put {1} [l] at -3.5 25
	\put {0} [b] at 5 4.5
	\put {1} [t] at -5 -4.5
	\setdashes
	\plot 32.5 2.5 5 2.5 5 -25 /
	\plot -32.5 -2.5 -5 -2.5 -5 25 /
	\setquadratic
	\plot 0 -35 5 -30.5 15 -22 20 -18 25 -14 30 -10.5 35 -7.5 40 -5 45 -2.5 50 -0.5 55 0
/
	\plot 0 35 -5 30.5 -15 22 -20 18 -25 14 -30 10.5 -35 7.5 -40 5 -45 2.5 -50 0.5 -55 0
/
\epc}

%ý same as before
%\midinsert
\vskip1cm \centerline{\figureFpex} \vskip1cm

\def\figureFdex{
\bpc
	\setcoordinatesystem units <0.6mm,0.6mm>
	\setplotarea x from -55 to 55, y from -55 to 55
	\setlinear
	\plot 0 55 -55 0 0 -55 55 0 0 55 /
	\plot 5 -25 32.5 2.5  0 35 /
	\plot -5 25 -32.5 -2.5  0 -35 /
	\plot 32.5 4.5 30 7 /
	\plot 55 2 52.5 4.5 /
	\multiput {$\bullet$} at 0 55 0 -55 55 0 -55 0 0 35 0 -35 32.5 2.5 -32.5 -2.5 5 -25 -5 25 5 2.5 -5 -2.5
/
	\put {0} [l] at 	57 0
	\put {1} [l] at 34.5 2.5
	\put {1} [r] at -57 0
	\put {1} [r] at -34.5 -2.5
	\put {$\lambda_1$} [b] at -18.75 -1.5
	\put {$\mu_2$} [b] at 18.75 3.5
	\put {$\nu_1$} [l] at -3.5 11.25
	\put {$\nu_2$} [r] at 3.5 -11.25
	\put {$\mu_1$} [b] at -25 16
	\put {$\lambda_2$} [t] at 25 -16
	\put {2} [b] at 0 57
	\put {0} [b] at 0 37
	\put {1} [t] at 0 -37
	\put {1} [t] at 0 -57
	\put {0} [r] at 3.5 -25
	\put {2} [l] at -3.5 25
	\put {0} [b] at 5 4.5
	\put {0} [t] at -5 -4.5
	\setdashes
	\plot 32.5 2.5 5 2.5 5 -25 /
	\plot -32.5 -2.5 -5 -2.5 -5 25 /
	\setquadratic
	\plot 0 -35 5 -30.5 15 -22 20 -18 25 -14 30 -10.5 35 -7.5 40 -5 45 -2.5 50 -0.5 55 0
/
	\plot 0 35 -5 30.5 -15 22 -20 18 -25 14 -30 10.5 -35 7.5 -40 5 -45 2.5 -50 0.5 -55 0
/
\epc}

%ý same as before
\centerline{\figureFdex}\vskip1cm
%\endinsert
 
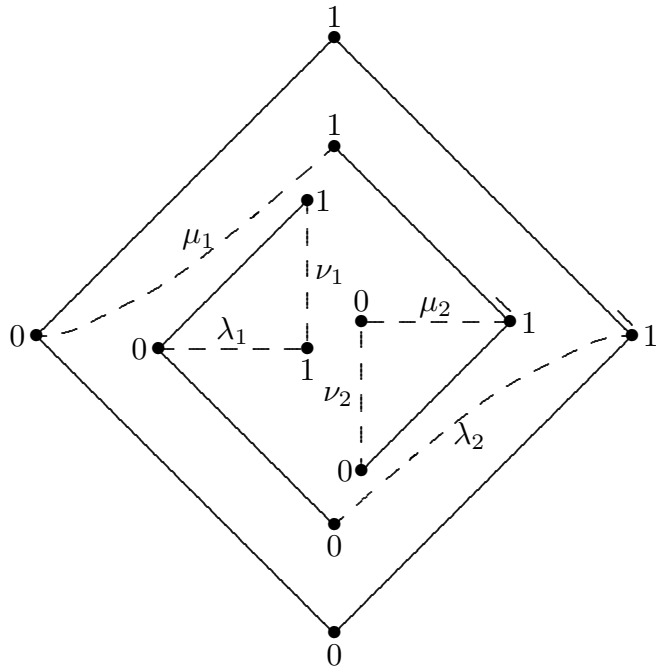
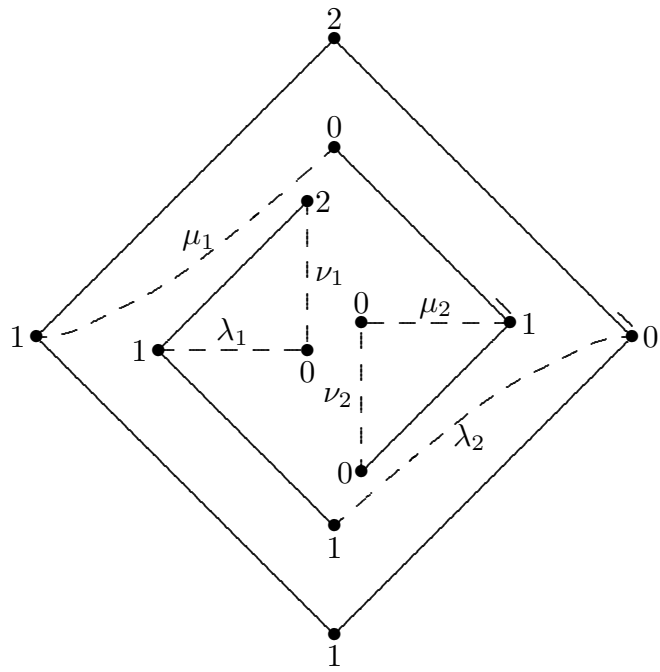
\centerline{Figure 2}
\vfill\eject
\end

$$\eqalign{ {G}&=\left(\prod_{i\not=1,3,5}~(1- E_i)^{-1}\right) \left(1+{ E_1
\over (1- E_1) (1- E_5)} \right.\cr & \left.\qquad\qquad +{ E_3  \over (1- E_3)
(1- E_1)}+{ E_5
\over (1- E_5) (1- E_3)}\right)\cr
&=\left(\prod_{i\not=2,4,6}~(1- E_i)^{-1}\right) \left(1+{ E_2 \over
(1- E_2) (1- E_4)} \right.\cr & \left.\qquad\qquad+ { E_4  \over (1- E_4)
(1- E_6)}+{ E_6 \over (1- E_6) (1- E_2)}\right)\cr
 &=\left(\prod_{i\not=7,8}~(1- E_i)^{-1}\right) \left(1+{ E_7
\over(1- E_7)}+{E_8 \over (1- E_8)}\right) \cr}\eqlabel\fgsutr$$